\documentclass[aps,
prd,
superscriptaddress,
onecolumn,
nofootinbib,
10pt]{revtex4-2}

\usepackage[colorlinks=true,
linkcolor=blue,
breaklinks=true,
urlcolor=magenta,
citecolor=blue]{hyperref}
\usepackage{graphicx,xcolor}% Include figure files
\usepackage{xtab,afterpage}
\usepackage{lipsum}
\usepackage{extarrows}
\usepackage{bm}% bold math
\usepackage{amssymb,amsmath,amsfonts, mathrsfs}
\usepackage{makecell}
\begin{document}
\title{Revisiting  $O(N)$ $\sigma$ model at unphysical pion masses and high temperatures}



\author{Yuan-Lin Lyu}
\email[]{yllyu@stu.pku.edu.cn}
\affiliation{School of Physics, Peking  University, Beijing  100871, People's Republic of China}

\author{Qu-Zhi Li}
\email[]{liquzhi@scu.edu.cn, corresponding author}
\affiliation{Institute for Particle and Nuclear Physics, College of Physics, Sichuan University, Chengdu  610065, People's Republic of China}

\author{Zhiguang Xiao}
\email[]{xiaozg@scu.edu.cn, corresponding author}
\affiliation{Institute for Particle and Nuclear Physics, College of Physics, Sichuan University, Chengdu  610065, People's Republic of China}

\author{Han-Qing Zheng}
\email[]{zhenghq@pku.edu.cn}
\affiliation{Institute for Particle and Nuclear Physics, College of Physics, Sichuan University, Chengdu  610065, People's Republic of China}

% \date{February 29, 2024}

\begin{abstract}
Roy-equation analysis on lattice data of $\pi\pi$ scattering phase
shifts at $m_\pi=391$ MeV reveals that the lowest $f_0$ meson becomes a
bound state under this condition. In addition, there is a pair of complex poles below
threshold generated by crossing symmetry~[X.-H. Cao \textit{et al.}, Phys. Rev. D \textbf{108}, 034009 (2023)]. We use
the $N/D$ method to partially recover crossing symmetry of  the $O(N)$
$\sigma$ model amplitude at leading order of $1/N$ expansion, and
qualitatively reproduce  the pole structure and pole trajectories with
varying pion masses as revealed by Roy-equation analyses. The $\sigma$
pole trajectory with varying temperature is also discussed and found to be similar to its properties when varying $m_\pi$. As the
temperature increases, the complex $\sigma$ poles firstly move from the
second Riemann sheet to the real axis becoming two virtual state
poles, and then one virtual state pole moves to the first sheet
turning into a bound state pole and finally tends to the pion pole position at high
temperature which is as expected from the chiral symmetry restoration.
Our results provide further evidences that the lowest $f_0$ state
extracted from experiments and lattice data plays the role of $\sigma$
meson in the spontaneous breaking of chiral symmetry.  Finally, we also briefly discuss the problems of the effective potential in the situation when $m_\pi$ and temperature get large.

\end{abstract}

\maketitle

\section {Introduction}
Chiral symmetry breaking plays an important role in the QCD low energy
dynamics. It is already well-known that due to the smallness of the
$u$ and $d$ quark masses,  QCD possesses an approximate
$SU(2)_L\times SU(2)_R$ chiral symmetry,  and it is also well accepted
that this symmetry is spontaneously broken by the nonzero $\langle 0|\bar q
q|0\rangle$ and three pseudo-Goldstone bosons are generated which are
identified as the $\pi$ mesons observed in the low energy hadron
scatterings. Historically, the famous linear sigma model firstly
proposed by Gell-Mann and L\'evy in 1960~\cite{Gell-Mann:1960mvl} could
provide an effective field theory description of this symmetry. In
this model, another scalar field $\sigma$ is combined with the three
pions to form a linear realization of an $O(4)$ symmetry and acquires a
vacuum expectation value (VEV) to break the $O(4)$ to $O(3)$, where the
$O(4)\simeq SU(2)_L\times SU(2)_R$ can be identified as the
previous chiral symmetry and the remaining $O(3)$ corresponds to the
preserved $SU(2)_V$. For a long time, the existence of the $\sigma$
particle was in controversy. The mild rise of the $\pi\pi$ phase shift
can hardly be recognized as generated from a typical resonance. A broad resonance
was proposed to describe the $\pi\pi$ scattering phase shift in 1960s,
see for example~\cite{Islam:1964zz, Patil:1964zz, Hagopian:1966zz}.
However, such a broad resonance  appeared and disappeared from the PDG
table several times from the 1960s until 2000s. Another description
using a nonlinear realization of the chiral
symmetry~\cite{Coleman:1969sm,Callan:1969sn} in which the
scalar-isoscalar particle is totally abandoned from the Lagrangian is
the nowadays very popular chiral perturbation
theory ($\chi$PT)~\cite{Gasser:1983yg,Gasser:1984gg}, which is regarded as
the low energy effective theory of QCD.  Within this
formalism, the low energy properties of the pion-pion scattering, such
as the scattering length, effective range, and phase shifts near the
threshold can be reproduced. The low energy coupling constants can be
saturated by integrating out vector
resonances~\cite{Ecker:1988te,Donoghue:1988ed}  (see however
\cite{Guo:2007ff,Guo:2007hm}). 
Thus, there seems to
be no need to have a scalar-isoscalar particle in describing the low
energy pion-pion scatterings. 
However, with energy going up, $\chi$PT
blows up quickly. Fortunately, unitarity and dispersive techniques come to its rescue. 
After unitarization, the $IJ=00$ channel $\pi\pi$
scattering amplitude dynamically generates a resonance state
represented as a pair of conjugate poles on the second Riemann sheet, 
see for example~\cite{Pelaez:2004xp,GomezNicola:2001as}.
However, this kind of unitarized method always generates more poles than physically
expected~\cite{Qin:2002hk}, especially spurious poles on the physical
sheet, which cast doubts on the reliability of the results, not to
mention the violation of crossing symmetry (for a recent review, see Ref.~\cite{Yao:2020bxx}). On the other hand, a novel
model-independent analysis representing the  partial wave $S$-matrix as a product of pole
and the left-hand cut integral terms, showed
that the left-hand cut estimated  from $\chi$PT always produces a negative
contribution to the phase shift while the data show a positive trend
near the threshold, which demonstrates the necessity of a subthreshold
resonance pole on the second sheet of the
amplitude~\cite{Xiao:2000kx}. This method was
further developed into the so-called Peking University (PKU)-representation of the partial wave $S$-matrix~\cite{Zheng:2003rw,Zhou:2004ms,Zheng:2003rv,Wang:2005ks,Zhou:2006wm}, and more precise mass and width for the particle was obtained
by fitting the data with the constraints from Balachandran-Nuyts-Roskies (BNR) relations derived from  
crossing symmetry~\cite{Zhou:2004ms}. A major step forward
 in analyzing the pion-pion scattering data is the so called
Roy-equation analysis~\cite{Roy:1971tc} which incorporates  crossing symmetry
and unitarity into a set of integral equations involving only the
partial wave amplitudes in the physical region. By solving these
integral equations with the low energy constraints from
$\chi$PT~\cite{Colangelo:2001df,Ananthanarayan:2000ht}  and extending
the solution to the complex plane, a broad resonance pole can be
found around $\sqrt{s_\text{pole}}=441^{+16}_{-8}-i272^{+9}_{-13}$ MeV~\cite{Caprini:2004zr}
(see also~\cite{Garcia-Martin:2011iqs} and~\cite{Mennessier:2008kk,Mennessier:2010xg}).
From these
model-independent efforts, the existence of a scalar-isoscalar
resonance in the low energy $\pi\pi$ scattering was firmly
established, and up to now PDG lists the particle as $f_0(500)$
with a pole mass in $400 \sim 550$ MeV and the half-width around $200\sim 350$ MeV.

Although these model-independent methods confirm the existence of this
particle, it is obscure  what the role of
this scalar-isoscalar particle plays in the low energy QCD spectrum
and chiral symmetry breaking.  An argument showing that this
$f_0(500)$ is not a usual pure $\bar qq$ state is that with large
number of colors, $N_c$, the pole generated in the unitarized $\chi$PT
goes away from the real axis which is different from the usual meson
behavior in the large $N_c$ limit~\cite{Pelaez:2004xp}. However, for
large enough $N_c$, the pole position still moves towards the real
axis~\cite{Sun:2005uk}. Another problem is whether it really
corresponds to the original $\sigma$ particle in the Lagrangian models
with  linearly realized chiral symmetry, 
which is still not clearly understood (see for example,
Ref.~\cite{Weinberg:2013cfa}).  Within the unitarized $\chi$PT
framework, the $f_0(500)$ particle is only a dynamically generated
resonance without any useful information about the role it plays in
the chiral symmetry breaking. This can be understood since  $\chi$PT
starts off from the broken phase of QCD with a nonzero VEV, it may not
provide much information about the global property of the effective
potential which is responsible for symmetry breakdown.
Moreover, a study of $\chi$PT at high temperatures
reveals its drastic difference compared with linear $\sigma$ model:
the former simply cannot restore the wanted $O(4)$ symmetry
explicitly, though there are some implicit evidences~
\cite{Gasser:1987ah,Gerber:1988tt,Cortes:2015emo,GomezNicola:2017bhm,
Gao:2019idb}. So to analyze the role played by this particle, it is
desirable to look back at the linear $\sigma$ model to see whether the
sigma particle in this model is consistent with $f_0(500)$ in 
model-independent analyses. One advantage of $O(N)$
linear $\sigma$ model is that in the large $N$ limit (here $N$ denotes
the number of flavors),  the model is exactly
solvable~\cite{Dolan:1973qd,Schnitzer:1974ji,Coleman:1974jh} and has been used to study the possible
relation between $f_0(500)$ and the $\sigma$~\cite{Guo:2006br}.

With the recent development of lattice QCD~\cite{Luscher:1986pf,Luscher:1990ck,Luscher:1990ux,Kuramashi:1993ka,He:2005ey,Mathur:2006bs,Feng:2009ij,Fu:2011xz,Briceno:2016mjc,HadronSpectrum:2008xlg,Rodas:2023gma}, 
phase shifts of $\pi\pi$ scattering can be reproduced from the first principle at various
unphysical pion masses.  This provides additional valuable information for the
resonances in  $\pi\pi$ scattering by extracting pole positions from those phase
shifts. In Ref.~\cite{Briceno:2016mjc}, the
scattering phase shifts for $IJ=00$ channel are calculated at
$m_\pi \sim 236$ and $391$ MeV, and
$K$-matrix parametrization was used to extract the poles in this
channel. The result shows that when $m_\pi \sim 236$ MeV the lowest $f_0$
state is still a resonance while at $391$ MeV it becomes a
bound state. 
More recent results by HadSpec collaboration using the similar $K$-matrix
method shows that 
at $m_\pi \sim 330$ MeV, $\sigma$ already becomes a shallow bound state,
whereas at $m_\pi \sim 283$ MeV it may become a virtual state or a
subthreshold resonance~\cite{Rodas:2023gma}.
In fact, it has been suggested that in the
unitarized $\chi$PT amplitudes with larger pion masses, the dynamically generated
$f_0$ particle moves towards the real axis below threshold and
finally becomes a bound state~\cite{Hanhart:2008mx,Pelaez:2010fj,Hanhart:2014ssa}.
However, it is well known that the $K$-matrix approach does
not satisfy crossing symmetry~\cite{Guo:2007ff,Guo:2007hm}, which is important in the low
energy pion-pion scattering and is crucial in determining the
properties of $f_0(500)$ resonance~\cite{Zhou:2004ms,Mennessier:2008kk,Mennessier:2010xg}. 
The first attempt to incorporate crossing symmetry in the analysis
is in Ref.~\cite{Gao:2022dln}, where the PKU
representation combined with
BNR relations and a virtual state accompanying the $f_0$
bound state was found at $m_\pi=391$ MeV. However, it was noticed that the
left-hand cut considered in that paper is not complete:  the
left-hand cut contribution introduced by the $f_0$ bound state from
crossed-channels  was not taken into account~\cite{Gao:2022tlh}. 
 Not only in Ref.~\cite{Gao:2022dln}, the  situation in general is also unsatisfactory. Model-independent lattice
data were always analysed using rough models.
A precise model-independent analysis of lattice data using Roy equation which
incorporates  crossing symmetry from the startup was done
in Ref.~\cite{Cao:2023ntr}. It is also found that on the second sheet there is a pair of conjugate
subthreshold poles  generated, which is related to the left-hand cut originated from
the $f_0$ bound state in crossed-channels. A general argument of
why this subthreshold pole is present was also given in Ref.~\cite{Cao:2023ntr}. 
Recently, a further lattice study using Roy-equation analysis at $m_\pi \sim 239$ and $283$ MeV found that for the former, the lowest $f_0$ state remains a resonance, whereas in the latter case the state was claimed (though not definitively, see Ref.~\cite{Rodas:2024qhn} for details) to become a virtual state, accompanied by a ``noisy" pole close to the left-hand cut on the second Riemann sheet. However, this additional virtual state pole is generated before the $f_0$ particle turns into a bound state, a situation not considered in Ref.~\cite{Cao:2023ntr}. To one's surprise, this phenomenon is exactly what happens in $N/D$ modified $O(N)$ model and we will explain the details in Sec.~\ref{sect:sigma}. 

The phenomena at unphysical pion masses provide a new
test ground of whether the lowest $f_0$ state in model-independent
analyses corresponds to the $\sigma$ particle in linear $\sigma$
model. The purpose of this paper is to use the solvable $O(N)$
linear $\sigma$ model with a crossing symmetry
improvement at unphysical pion masses and compare it with the results
from  Roy-equation analyses. 

The paper is organized as follows. In Sec.~\ref{sect:LO}, a review on standard results of $O(N)$ model is given,  especially the behavior of
the lowest $f_0$ particle is discussed and compared with similar results obtained by different model analyses of the lattice data. The comparison necessitates the effort of going beyond the lowest order calculation of $O(N)$ model.  
In Sec.~\ref{sect:sigma}, the modified $O(N)$ amplitude incorporating crossing symmetry is introduced, 
aided by the use of $N/D$ method. The  results from Roy-equation analyses on the $\sigma$ particle can be
reproduced  in this approach  at a qualitative level. In particular, the
$f_0$ state becomes a bound state and the lower subthreshold pole generated through
cross-channel effects also emerges at $m_\pi=391$ MeV after partially imposing
crossing symmetry, by tuning the parameters
in $N/D$ method.
 Though it is well-known that the linear $\sigma$ model is not QCD~\cite{Gasser:1983yg,Gasser:1984gg},  this paper is trying to demonstrate that it provides a qualitative description of low energy QCD in $IJ=00$ channel at phenomenological level, even for unphysical pion masses -- from this observation we take the perspective that $f_0(500)$ particle plays the role of the $\sigma$ particle. In Sec.~\ref{sect:temperature},  we also investigate thermal properties of the scattering amplitudes with leading order $1/N$ expansion. We reproduce the widely accepted results that $O(N)$ symmetry is restored at high temperature, irrespective of different $m_\pi$ values. Finally Sec.~\ref{sect:conclusion} is for discussions and conclusions,  where  we shortly discuss the problem that the effective potential no longer provides a local minimum at high temperature, as well as when $m_\pi$ gets large. Future improvements on the related issues are also outlined there.

\section {$\sigma$-pole  with varying $m_\pi$ in $O(N)$ model \label{sect:LO}}

In this section, we will review the $O(N)$ linear $\sigma$ model and
look at  the $\sigma$-pole trajectory with varying $m_\pi$, at leading order of $1/N$ expansion. The
Lagrangian for this model is
\begin{align}
	\mathcal L_{O(N)} = \frac{1}{2}\partial_\mu \phi_a \partial^\mu \phi_a -\frac{1}{2}\mu_0^2 \phi_a \phi_a - 
	\frac{\lambda_0}{8N}(\phi_a\phi_a)^2+ \alpha \phi_N \,,
\end{align}
where $a=1,2,\cdots,N$. When $\mu_0^2<0$, without the linear $\alpha$
term, the system has a
spontaneous symmetry breaking of $O(N)\to O(N-1)$, $\langle \phi\rangle\neq 0$. With the
linear $\phi_N$ term, the VEV is aligned with the $N$th direction,
 namely, $\langle \phi_N\rangle =v$,  with $N-1$ pseudo-Goldstone particles,
$\pi_a\equiv\phi_a$, $a=1,\dots,N-1$ and we define the shifted field $\sigma \equiv \phi_N- v$. 
For convenience when counting 
$1/N$ orders in the calculation of the effective action, we introduce an auxiliary field
$\chi$ to the Lagrangian~\cite{Coleman:1974jh}, 
\begin{align}
	\mathcal L \to \mathcal L + \frac{N}{2\lambda_0}\left(\chi-\frac{\lambda_0}{2N}\phi_a\phi_a -\mu_0^2 \right)^2
	=  \frac{1}{2}\partial_\mu \phi_a \partial^\mu \phi_a +\alpha\phi_N +\frac{N}{2\lambda_0} \chi^2 -\frac{1}{2}\chi\phi_a\phi_a -\frac{N\mu_0^2}{\lambda_0}\chi \,,
\end{align}
with an irrelevant constant omitted. The effective action can be
obtained by standard procedures,
\begin{align}
	\Gamma(\phi,\chi)=& \int \mathrm d^4 x \left( \frac{1}{2}\partial_\mu \phi_a \partial^\mu \phi_a +\alpha\phi_N +\frac{N}{2\lambda_0} \chi^2 -\frac{1}{2}\chi\phi_a\phi_a -\frac{N\mu_0^2}{\lambda_0}\chi   \right) \notag\\
	&+\frac{i}{2} N \mathrm{Tr} \log ( \partial^2 +\chi - i\epsilon ) \,,
\end{align}
where $\mathrm{Tr}$ denotes trace taken in 4-dimensional Minkowski spacetime and $\epsilon \to 0^+$. For convenience we will not distinguish the notation for the  classical fields and the original fields in the Lagrangian, which can be understood in the context.
Then the effective potential can be obtained as 
\begin{align}
	V(\phi,\chi) = -\alpha\phi_N -\frac{N}{2\lambda_0} \chi^2 +\frac{1}{2}\chi\phi_a\phi_a +\frac{N\mu_0^2}{\lambda_0}\chi   
	-\frac{i}{2} N \int \frac{\mathrm{d}^4 \ell}{(2\pi)^4} \log ( -\ell^2 +\chi -
i\epsilon ) \,,
\end{align}
where $\phi_a$, $\chi$ are regarded as their constant expectation
values, respectively.
The
renormalization conditions are chosen to be~\cite{Coleman:1974jh,Chivukula:1991bx}
\begin{align}
		\frac{\mu(M)^2}{\lambda(M)}&=\frac{\mu_0^2}{\lambda_0}
+\frac{i}{2}  \int \frac{\mathrm{d}^4 \ell}{(2\pi)^4} \frac{1}{ \ell^2 + i\epsilon}
\,,
\label{eq:ren-cond1}\\
		\frac{1}{\lambda(M)} &=\frac{1}{\lambda_0}
-\frac{i}{2}  \int \frac{\mathrm{d}^4 \ell}{(2\pi)^4} \frac{1}{ (\ell^2  + i\epsilon)
(\ell^2 -M^2 + i\epsilon)}\,.
\label{eq:ren-cond2}
\end{align}
With these conditions, the minimum of the effective potential can be
obtained by solving $\frac{\partial V}{\partial \chi} = 0$ and
$\frac{\partial V}{\partial \phi_a} = 0$ which reduce to
relations of the VEVs for the corresponding
fields
\begin{gather}
	\phi_a\phi_a =\frac{2N}{\lambda}\chi - \frac{2N\mu^2}{\lambda}
- \frac{N}{16\pi^2}\chi \log \frac{\chi}{M^2}\,,
\label{eq:zeroT Gap eq 1}\\
	\chi \phi_a =0 \ (a<N), \quad \chi\phi_N -\alpha=0\,.
\label{eq:zeroT Gap eq 2}
\end{gather}
Thus, the VEV for each $\pi_a$ is still zero, i.e.
$\langle \pi_a\rangle=0$ and $\phi_N$ gets expectation value $\langle
\phi_N\rangle =v=\alpha/\langle \chi\rangle$.
Since the correction to the pion
mass term  only appears at higher $1/N$ order, at leading order we have $\langle \chi\rangle =m_\pi^2$.
At zero-temperature, from the definition of the pion decay constant $f_\pi$, $\langle
0|A_a^\mu(x)|\pi\rangle=i p^\mu f_\pi e^{-ip\cdot x}$, where $A_a^\mu$ is the axial vector current, and the relation of partially conserved axial current (PCAC), $\partial_\mu A_a^\mu=\alpha \pi_a$, we have $v=f_\pi$. 
For phenomenological calculations, we always set $f_\pi=92.4\,$MeV and $N=4$.

With these preparations, now the $\pi\pi$ scattering amplitude in
the leading  $1/N$ order can be expressed according to the external
isospin structure as follows
\begin{align}
	\mathcal T_{\pi_a\pi_b\to \pi_c\pi_d} = i D_{\tau\tau}(s) \delta_{ab}\delta_{cd} 
	+i D_{\tau\tau}(t) \delta_{ac}\delta_{bd}  
	+i D_{\tau\tau}(u) \delta_{ad}\delta_{bc} \,,
\label{eq:T}
\end{align}
with $D_{\tau\tau}$ the propagator for $\tau\equiv\chi-\langle \chi\rangle$ field from the
effective action. Since there are mixing terms of $\tau$ and $
\sigma$, a $2\times 2$ inverse propagator matrix needs to be
considered~
\cite{Coleman:1974jh,Chivukula:1991bx}, which can be expressed  in  the momentum space as
\begin{align}\label{eq: Inverse sigma tau propagator at LO}
	D^{-1}(p^2)= -i
	\begin{pmatrix}
		p^2 -m_\pi^2 & -f_\pi \\
		-f_\pi   &        N/ \lambda_0 + N B_0(p^2,m_\pi)
	\end{pmatrix}\,,
\end{align}
with 
\begin{align}\label{eq:def of B0 function}
	B_0(p^2,m_\pi) = \frac{-i}{2}\int \frac{\mathrm{d}^4 \ell}{(2\pi)^4}
\frac{1}{ (\ell^2 -m_\pi^2 + i\epsilon)  ((\ell+p)^2 -m_\pi^2 +
i\epsilon)}\,.
\end{align}
The propagators can then be obtained as
\begin{align}
	D_{\tau\tau}(p^2) &= \frac{i (p^2 -m_\pi^2) }{ (p^2 -m_\pi^2)(N/ \lambda_0 + N B_0(p^2,m_\pi)) -f_\pi^2}, \\
	D_{\sigma\sigma}(p^2) &= \frac{i (1/ \lambda_0 +
B_0(p^2,m_\pi)) }{ (p^2 -m_\pi^2)(1/ \lambda_0 +  B_0(p^2,m_\pi))
-f_\pi^2/N}\,.
\end{align}
Using the previous renormalization conditions
Eqs.(\ref{eq:ren-cond1},\ref{eq:ren-cond2}),
we have 
\begin{align}
	\frac{1}{\lambda_0}+ B_0(p^2,m_\pi) &=\frac{1}{\lambda(M)}+
B(p^2,m_\pi,M)\,,
\\ \label{eq: def of ren-Bfun}
	B(s,m_\pi,M) &= \frac{1}{32\pi^2} \left( 1 + \rho(s)
\log\frac{\rho(s)-1}{\rho(s)+1}  -\log\frac{m_\pi^2}{M^2}\right)\, ,
\end{align}
where $\rho(s) = \sqrt{1-4m_\pi^2/s}$. For our purpose, we need only  $IJ=00$ amplitude in the leading
order of $1/N$ expansion, 
\begin{align}
	\mathcal T^{LO}_{00}(s)=\frac{i N D_{\tau\tau} (s)}{32\pi}\,,
\end{align}
which is of $\mathcal O(1)$.

The
$\sigma$ resonance corresponds to poles on the second Riemann
sheet, which can be obtained by solving the zero points of the denominator of $D_{\tau\tau}$:
\begin{equation}
	 (s -m_\pi^2)(1/ \lambda(M) + B^{II}(s,m_\pi,M)) -f_\pi^2/N=0 \,,
\label{eq:pole-pos-eq}
\end{equation}
where $B^{II}$ represents the analytically continued
$B(s,m_\pi,M)$ function onto the second sheet, which is obtainable by changing the sign of $\rho(s)$.

Since the coupling constant $\lambda$  and the renormalization scale
$M$ are related, we can define $M$ to be the intrinsic scale of $O(N)$ model, when regarded as an
effective field theory, at which the coupling blows up, i.e.
$1/\lambda(M)=0$~\cite{Chivukula:1991bx}. Then the coupling $\lambda$ would not appear in the
scattering amplitude:
\begin{align}
	\mathcal T^{LO}_{00}(s)= -\frac{1}{32\pi} \frac{s -m_\pi^2 }{ (s -m_\pi^2)
B(s,m_\pi,M) -f_\pi^2/N}\,,
\label{eq:LO-T00}
\end{align}
and Eq.(\ref{eq:pole-pos-eq}) can be recast into
\begin{equation}
		 (s -m_\pi^2)B^{II}(s,m_\pi,M) -f_\pi^2/N=0\,.
\label{eq:pole-eq}
\end{equation}
The leading order amplitude automatically satisfies the exact partial wave
unitarity, i.e. $\mathrm {Im}\, \mathcal T_{00}(s) =\rho(s) |\mathcal T_{00}(s)|^2$.
It also has an Adler zero~\cite{Adler:1964um} at $s=m_\pi^2$.  
Notice that the isospin projection for $O(N)$ singlet channel is
$\mathcal T_{I=0}(s,t,u) = (N-1) i D_{\tau\tau}(s) + i D_{\tau\tau}(t) + i D_{\tau\tau}(u)$.
Thus the $t$- and $u$-channel amplitudes in Eq.~(\ref{eq:T})
contribute to the $\mathcal O(1/N)$ amplitude, and only the $s$-channel
amplitude contributes to the leading order $IJ=00$ partial wave  amplitude. This already breaks  crossing
symmetry.

\begin{figure}
\includegraphics[width=0.5\textwidth]{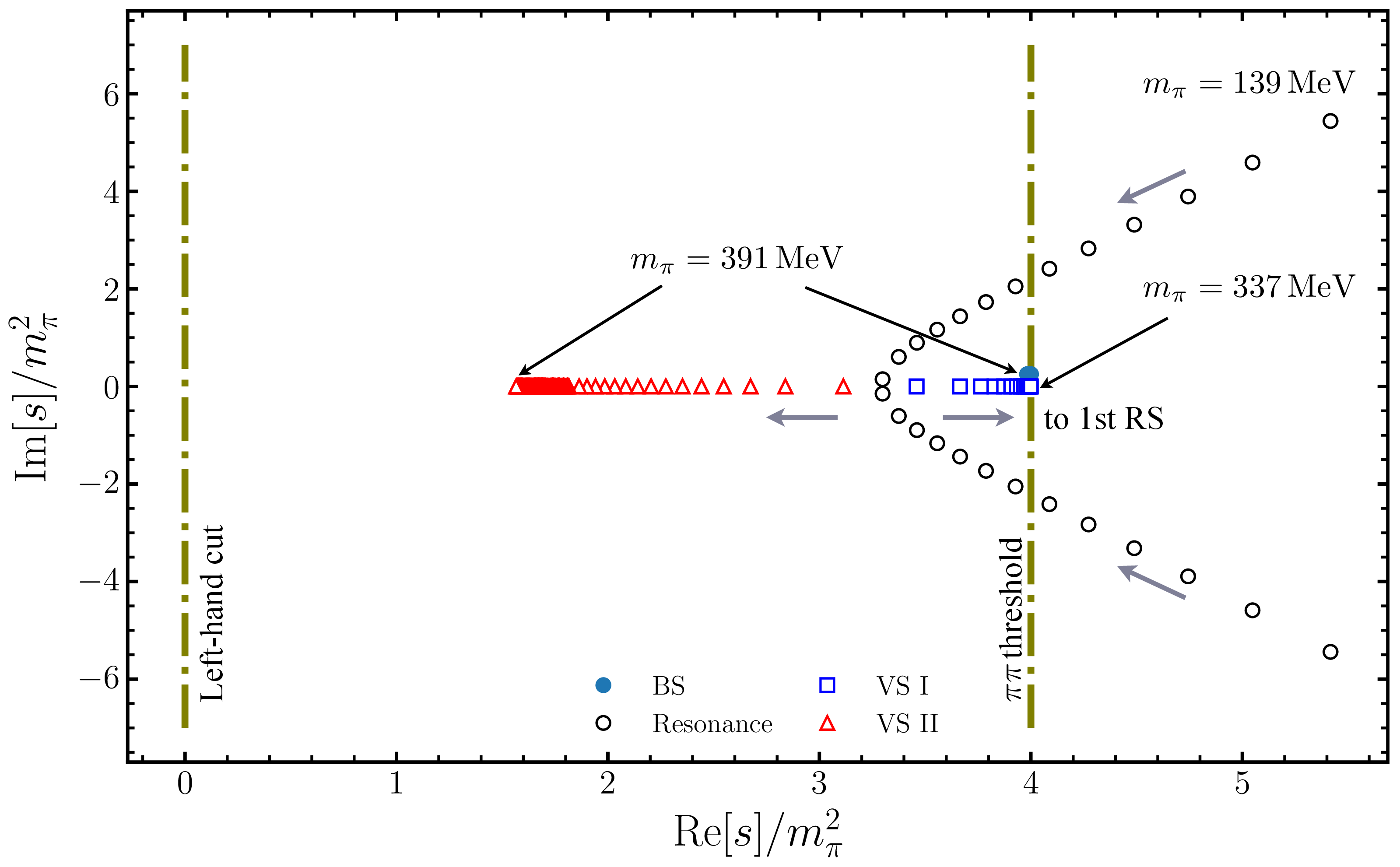}
\caption{ The pole trajectories for the $\sigma$ poles of the leading
order $O(N)$ amplitude. The left-hand cut branch point is still at $s_L =0$ after the virtual state pole VS I moves across threshold to the physical Riemann sheet (RS), becoming a bound state. There is neither additional virtual state pole generated from the left-hand cut (suggested by Refs.~\cite{Cao:2023ntr,Rodas:2024qhn}), nor subthreshold poles similar to those found in Ref.~\cite{Cao:2023ntr}. \label{fig:sigma-traj-ON}}
\end{figure}

With the leading order $\mathcal T$-matrix, we can  study the $\sigma$ pole trajectory with varying $m_\pi$. The intrinsic scale $M$ is chosen at $550\,\mathrm{MeV}$. See Table~\ref{tab:poles} for comparison of the pole positions obtained in  $O(N)$ model and
the lattice results analyzed using $K$-matrix and Roy equation respectively.
The pole trajectory is shown in Fig.~\ref{fig:sigma-traj-ON}. When 
$m_\pi$ increases  from the physical mass, the
$\sigma $ poles   move towards the real axis, and then become two
virtual state poles (VS I and II) after they meet at the real axis. One virtual
state pole (VS II) moves down away and the other (VS I) moves
towards the threshold, crossing it to the first sheet, and becomes a
bound state pole (BS).

\begin{table}
\begingroup
\setlength{\tabcolsep}{8pt} 
\renewcommand{\arraystretch}{1.5} 
\begin{tabular}{cccccc}
\hline
\hline
$m_\pi$ (MeV)&$139$
&$239$
&$283$&$330$&$391$
\\
\hline\vspace{-8pt}\\
\vspace{5pt}
$O(N)$ (LO)
&$356 - i 148$ 
&$448 - i 57$
&$\begin{gathered}
558(\text{VS I}) \\ 438(\text{VS II})
\end{gathered}$
&$\begin{gathered}
660(\text{VS I}) \\ 451(\text{VS II})
\end{gathered}$
&$\begin{gathered}
780(\text{BS}) \\ 489(\text{VS II})
\end{gathered}$
\\
\hline\vspace{-8pt}\\
\vspace{5pt}
$N/D$ modified $O(N)$ 
&$348 - i 180$ 
&$\begin{gathered}
469(\text{BS}) \\ 426(\text{VS II}) \\ 168(\text{VS III})
\end{gathered}$
&$\begin{gathered}
527(\text{BS}) \\ 422(\text{VS II}) \\ 264(\text{VS III})
\end{gathered}$
&$\begin{gathered}
585(\text{BS}) \\ 396 - i 28 \\ (\text{Sub. pole})
\end{gathered}$
&$\begin{gathered}
658(\text{BS}) \\ 466 - i 77 \\ (\text{Sub. pole})
\end{gathered}$
\\
\hline\vspace{-8pt}\\
\vspace{5pt}
lattice + $K$-matrix & 
&\makecell[l]{
    $(487 \sim 809$ \\
    $- i 136 \sim 304)$\cite{Briceno:2016mjc,Rodas:2023gma}}
&\makecell[l]{
    $(476 \sim 579$ \\
    $- i 0 \sim 129)$\cite{Rodas:2023gma}}
& $657^{+3}_{-4}(\text{BS})$\cite{Rodas:2023gma}
&$758\pm 4(\text{BS})$\cite{Briceno:2016mjc}
\\
\hline\vspace{-8pt}\\
\vspace{5pt}
lattice + Roy Eq. & 
&\makecell[l]{
    $(416 \sim 644$ \\
    $- i 176 \sim 307)$\cite{Cao:2023ntr,Rodas:2024qhn}}
&\makecell{
    $522\sim 562$ \\ 
    (VS I\&II)\cite{Rodas:2024qhn} \footnote{ Additionally, there is a third though ``noisy" pole close to the left-hand cut on the second sheet, which could correspond to the virtual state pole VS III appearing in the $N/D$ modified $O(N)$ model analysis. However, a definitive conclusion about whether the $\sigma$ is a virtual state or a subthreshold resonance at this $m_\pi$ value cannot be made, owing to large statistical uncertainties in the results, see Ref.~\cite{Rodas:2024qhn} for details.} }
&
&\makecell{
    $759^{+7}_{-16}(\text{BS})$\cite{Cao:2023ntr}\\
    $269^{+40}_{-25} -i 211^{+26}_{-23}$\\
    (\text{Sub. pole})\cite{Cao:2023ntr}}

\\ 
\hline
\hline
\end{tabular}
\endgroup
\caption{\label{tab:poles} Comparison of the pole positions ($\sqrt{s_\text{pole}}$) for $O(N)$
model, lattice + $K$-matrix~\cite{Briceno:2016mjc,Rodas:2023gma}
 and
lattice + Roy equation~\cite{Cao:2023ntr,Rodas:2024qhn}. When $m_\pi = 391$ MeV, the subthreshold (Sub.) pole close to the left-hand cut in Ref.~\cite{Cao:2023ntr} can also be found (qualitatively) within the $N/D$ modified $O(N)$ model discussed in this work.
}
\end{table}

From Eq.~\eqref{eq:pole-eq} one can work out the condition for the
critical pion mass $m_\pi=m_{c}$ at which the
$\sigma$ pole is located exactly at the threshold,
\begin{equation}
\log\frac{m_{c}^2}{M^2}=1-\frac{32\pi^2f_\pi^2}{3 m_{c}^2
N}\,.
\end{equation}
The numerical result is $m_{c}\simeq 337\, \mathrm{MeV}$.
This is a little bit different from the lattice results
which may be located at somewhere between $283$ and $330\, \mathrm{MeV}$~\cite{Briceno:2016mjc,Rodas:2023gma,Rodas:2024qhn}. The difference is not surprising since the $m_c$ obtained here is only
the leading $1/N$ order result of $O(N)$ model.
When $m_\pi>m_{c}$, the $\sigma$ particle becomes a bound state. At the same
time, it also appears in the $t$- and $u$-channel amplitudes due to
crossing symmetry, which are not included in the leading order
amplitude since they are of $\mathcal O(1/N)$ after isospin projection.
If we add the $t$- and $u$-channel
contributions and do the isospin and
partial wave projection, then the $\mathcal T_{00}$ matrix is expressed as 
\begin{align}\label{eq:ONamp cross channel}
	\mathcal T_{00}(s) = \frac{N-1}{32\pi} \mathcal A^{LO}(s) + I_{tu}(s) \,,
\end{align}
where
\begin{align}
	\mathcal A^{LO}(s) &=  \frac{m_\pi^2 - s}{ (s - m_\pi^2) N B(s,m_\pi,M)
- f_\pi^2}\,, \\
	I_{tu}(s) &= \frac{1}{16\pi (s - 4m_\pi^2)} \int_{4m_\pi^2 -
s}^0 \mathrm d t \mathcal A^{LO}(t)\,.
	\label{eq:ONamp Itu}
\end{align}
Function $I_{tu}(s)$ in Eq.~\eqref{eq:ONamp cross channel} comes from the partial
wave projection of the crossed $t$- and $u$-channel amplitudes. 
The bound state in the crossed-channels also generates a left-hand cut with a branch point at $4m_\pi^2-m_\sigma^2$.  
However, this amplitude does not satisfy the exact unitarity any more. 
Thus, to partially recover crossing symmetry and restore unitarity, we need to
resort to some unitarization methods, e.g. inverse amplitude method (IAM, see Ref.~\cite{GomezNicola:2007qj} and references therein) and $N/D$ method~\cite{Chew:1960iv}. The
situation here is not suitable for direct application of IAM,
because the partial wave amplitude Eq.~\eqref{eq:ONamp cross channel} is
not a complete calculation at $\mathcal O(1/N)$ and thus will break even the
perturbative version of unitarity relation, namely
\begin{align}\label{eq:perturbative unitarity for IAM}
\operatorname{Im}
\mathcal T^{NLO}_{00} = \rho \left[\mathcal T^{LO}_{00}(\mathcal
T^{NLO}_{00})^* +(\mathcal T^{LO}_{00})^* \mathcal T^{NLO}_{00}
\right] = \frac{2 \rho \operatorname{Re} \mathcal T^{LO}_{00} \operatorname{Re} \mathcal T^{NLO}_{00}}{1-2 \rho  \operatorname{Im} \mathcal T^{LO}_{00}}\,,
\end{align}
where we also represent $\operatorname{Im}
\mathcal T^{NLO}_{00}$, i.e. the imaginary part of the next-to-leading order amplitude, in terms of $\operatorname{Im}
\mathcal T^{LO}_{00}$ and $\operatorname{Re}
\mathcal T^{NLO}_{00}$ after simple algebraic calculations.
Nevertheless, it is still possible to acquire an approximation of the $\mathcal O(1/N)$ partial
wave (denoted by $\mathcal T^{NLO\prime}_{00}$) that does satisfy the above relation by setting $\operatorname{Re} \mathcal T^{NLO}_{00} = - \frac{1}{N} \operatorname{Re} \mathcal T^{LO}_{00} + I_{tu}$ and obtaining $\operatorname{Im}
\mathcal T^{NLO}_{00}$ using Eq.~\eqref{eq:perturbative unitarity for IAM}. After the analytic continuation of the modified amplitude $\mathcal T^{NLO\prime}_{00}$, the approximated ``IAM unitarized amplitude" $\mathcal T^\text{IAM}_{00}$ can be constructed as usual:
\begin{align}
    \mathcal T^{\text{IAM}}_{00}= \frac{\left(\mathcal T^{LO}_{00}\right)^2}{\mathcal T^{LO}_{00}-\mathcal T^{NLO\prime}_{00}}\,.
\end{align}
However, this unitarized amplitude is found to have spurious poles both on the 1st and 2nd sheets that may not be simply removed. Thus in the
following we will use the $N/D$ method.

\section {$\sigma$ pole in $N/D$ modified $O(N)$ model with varying
$m_\pi$ \label{sect:sigma}}

It has been seen that in $O(N)$ model the isospin projection to
$I=0$ channel breaks crossing symmetry at the leading order of $1/N$
expansion.
The $O(N)$ amplitude must be modified in order to at least partially recover crossing symmetry while preserving 
unitarity.  The key point is to generate the left-hand cut from crossed-channels in the
scattering amplitude, which should be consistent with the bound state  the
$\sigma$ generated. It is well known that the $N/D$ method can
introduce the left-hand cut contribution to the amplitude and at the
same time preserve  partial wave unitarity, which  is suitable for
the purposes here.

In the spirit of $N/D$ method, the $\mathcal T$-matrix
can be expressed as 
\begin{align}
	\mathcal T(s) = \frac{N(s)}{D(s)}\,,
\end{align}
where the singularity of $N(s)$ contains only the left-hand cut ($L$), while 
$D(s)$ only contains right-hand cut ($R$) and Castillejo-Dalitz-Dyson (CDD) poles~\cite{Castillejo:1955ed}, thus
$\operatorname{Im}_R N(s) = \operatorname{Im}_L D(s) =0 $.  To satisfy
the partial wave  unitarity relation, i.e. $\operatorname{Im}_R \mathcal T^{-1} = -
\rho$, the relation between $N(s)$ and $D(s)$ should be
\begin{align}\label{eq:ImR D in N/D}
	\operatorname{Im}_R D(s) &= -\rho(s)N(s)\,, \\
	\operatorname{Im}_L N(s) &= D(s)\operatorname{Im}_L \mathcal
T(s)\,.
\end{align}
Using the Cauchy integral formula, one can write down dispersion
relations for $N(s)$ and $D(s)$, and then solve $N(s)$ and $D(s)$
numerically to obtain the scattering amplitude.

Our strategy is to extract the $\mathrm{Im}_L \mathcal T$ from  $O(N)$
model, use $N/D$ method to obtain the scattering amplitude,
which  at the leading $1/N$ order recovers the original $O(N)$ model
amplitude, and require the position of the $\sigma$ bound state to be
consistent with the left-hand cut branch point 
at $s_L=4m_\pi^2-m_\sigma^2$ for large unphysical $m_\pi$.
Since the numerator and denominator of the amplitude
in Eq.~\eqref{eq:LO-T00} both have an $s$ power less than $s^2$ as $s\to
\infty$, it is natural to
use \textit{twice subtracted} dispersion
relations for $N(s)$ and $D(s)$ without CDD poles, and as a bonus,
the zero point of the $\mathcal T$-matrix 
can be  dynamically generated, which could correspond to the Adler zero.  The twice subtracted dispersion
relations for $N(s)$ and $D(s)$ can be expressed as
\begin{align}\label{eq: N/D twice sub for D}
		D(s) &= \frac{s-s_A}{s_0-s_A}  + g_D\frac{s-s_0}{s_A-s_0} - \frac{(s-s_0)(s-s_A)}{\pi} \int_R \frac{\rho(s')N(s')}{(s'-s)(s'-s_0)(s'-s_A)} 
		\mathrm d s' \,,\\
		N(s) &=b_0 \frac{s-s_A}{s_0-s_A} + g_N\frac{s-s_0}{s_A-s_0} + \frac{(s-s_0)(s-s_A)}{\pi} \int_L \frac{D(s')\mathrm{Im}_L \mathcal T(s')}{(s'-s)(s'-s_0)
		(s'-s_A)} \mathrm d s' 
		\label{eq: N/D twice sub for N}\,,
\end{align}
where the subtraction points $s_A = m_\pi^2$, $s_0 = s_{th}=4m_\pi^2$,
and $D(s_0)=1$ are chosen for convenience. With these choices, the
subtraction constants are
$N(s_0) = b_0$, $N(s_A) = g_N$, $D(s_A) = g_D$. Noticing that at
$s_A=m_\pi^2$, the leading order partial wave amplitude in Eq.~\eqref{eq:LO-T00} is zero, corresponding to the leading order Adler zero. Then with 
$\mathcal T(s_A)=N(s_A)/D(s_A)\sim \mathcal O(1/N)$, one can choose  $g_N$ to be $\mathcal O(1/N)$ and $g_D$ to be $\mathcal O(1)$. 
Thus to the leading order, $N(s)=b_0 \frac{s-s_A}{s_0-s_A}
$. By substituting this into Eq.~\eqref{eq: N/D twice sub for D}
one  recovers the leading order scattering amplitude by choosing  
\begin{align}
	b_0 &
=  - \frac{1}{32\pi}\frac
{s_0-s_A}{(s_0-s_A)B(s_0,m_\pi,M)-f_\pi^2/N}
\,, \label{eq:Constraint1 LO}
\\
g_D&=\frac{32\pi f_\pi^2 b_0}{N(s_0-s_A)}
\,. \label{eq:Constraint2}
\end{align}
It is obvious that  $b_0$ is just the leading order amplitude which is
consistent with our prescription  $N(s_0) = b_0$ and $D(s_0)=1$. After
including  $t$- and $u$-channel contributions in the $N/D$
construction, the amplitude recovers unitarity and  has a
left-hand cut with a branch point at $4m_\pi^2-m_\sigma^2$ when
$m_\pi>m_{c}$, which is as expected.  However, since the
$\sigma$ pole in the crossed-channels is still located at the same position in
the leading order amplitude, the branch point at $4m_\pi^2-m_\sigma^2$ is determined by the leading order $m_{\sigma}$ in $1/N$ expansion. But with the $\mathcal O(1/N)$ left-hand cut
contribution in $D(s)$, the mass of the $\sigma$ bound state may be shifted from the pole position of 
$\mathcal T^{LO}_{00}$. 
To be consistent with the left-hand cut, demanded by crossing symmetry, we need to
require the $\sigma$ bound state pole position (solved from $D(s)=0$) to be the
same as the one in leading order amplitude. In general, this can be
achieved by properly choosing the higher order corrections to the parameters
$b_0$ and $g_D$. For $m_\pi>m_{c}$ this can only be done numerically, our prescription is as follows: 
\begin{align}
	b_0 &= \mathcal T_{00}(s_0) = \frac{N-1}{32\pi} \mathcal A^{LO}(s_0) +
I_{tu}(s_0)\,, \label{eq:Constraint1}\\
    D(m_\sigma^2) &=0\,, \label{eq:Constraint2 Ssigma }\\
    \frac{g_N}{g_D} &= \operatorname{Re} I_{tu}(s_A)\,. \label{eq:Constraint3}
\end{align}
where $b_0$ is chosen as the amplitude evaluated at $s_0$ with the
crossed-channels added, $g_D$ is determined by requiring the bound state pole position to be the same as the leading order result $s_\sigma =m_\sigma^2$, and Eq.~\eqref{eq:Constraint3} results directly from the requirement $\mathcal T(s_A) = \mathcal T_{00} (s_A)$ while $s_L<s_A$\footnote{If $m_\sigma^2 < 3 m_\pi^2$ such that $s_A = m_\pi^2$ is located in the left-hand cut region, then $\mathcal T_{00} (s_A)$ is replaced by $\operatorname{Re} \mathcal T_{00} (s_A)$.}. When $m_\pi<m_{c}$, Eq.~\eqref{eq:Constraint2 Ssigma } is replaced by Eq.~\eqref{eq:Constraint2} with Eq.~\eqref{eq:Constraint1} and Eq.~\eqref{eq:Constraint3} not changed. 
This prescription for the subtraction constants captures the most important features of the inputted $O(N)$ model. Anyway, it is of course not the ``unique solution" for $N/D$ modified $O(N)$ model. We also tried several different sets of subtraction constants, and found that the qualitative results for the $\sigma$ pole trajectory are quite robust.

\begin{figure}
\includegraphics[width=0.5\textwidth]{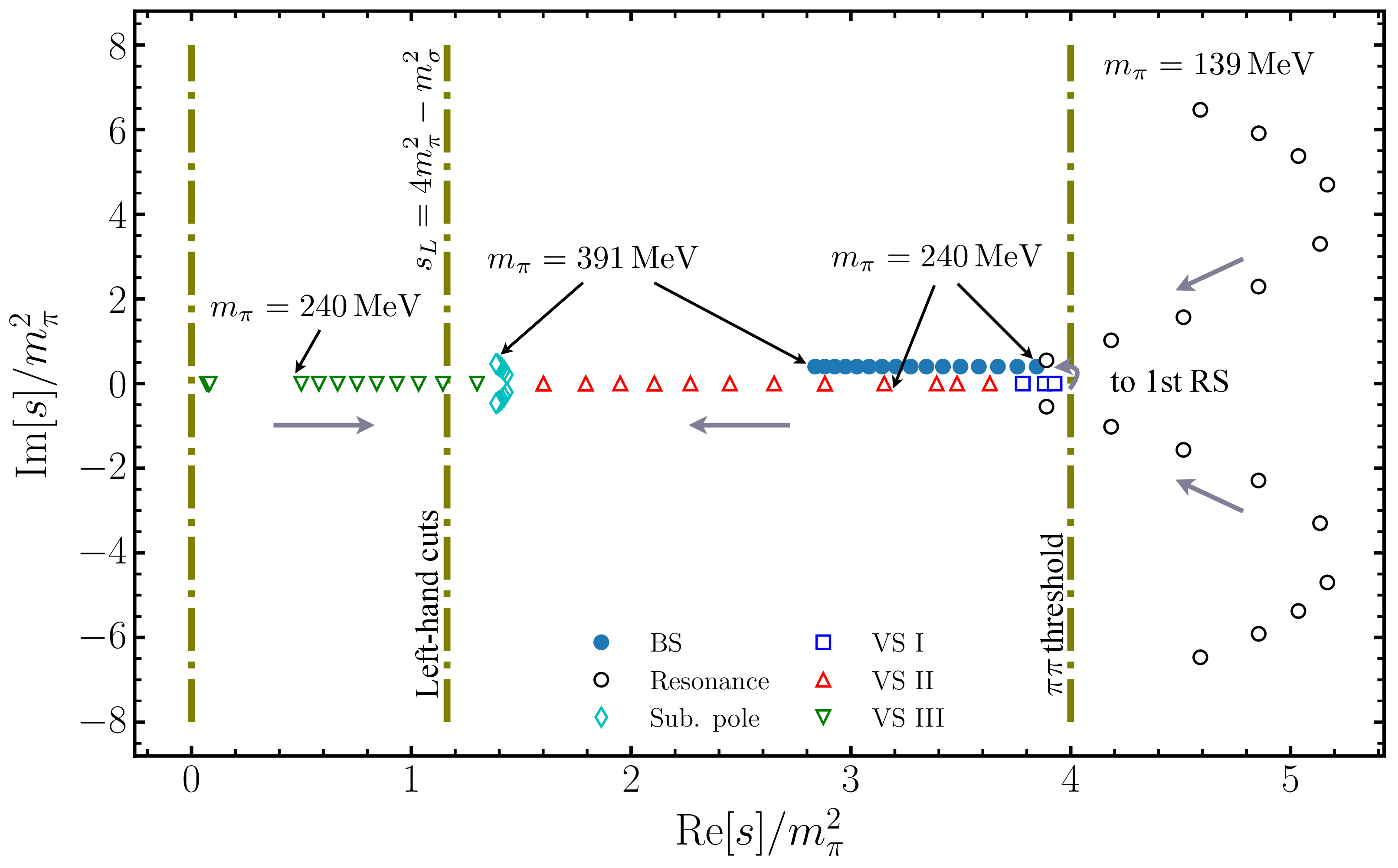}
\caption{
\label{fig:poletraj-ND}
The pole trajectories for $N/D$ modified $O(N)$ model. The left-hand cut branch point extends to $s_L = 4m_\pi^2 - m_\sigma^2$ after the virtual state pole VS I moves across threshold to the physical Riemann sheet (RS), turning into a bound state. Qualitatively, the $N/D$ modified $O(N)$ model reproduces the picture derived from $K$-matrix~\cite{Briceno:2016mjc,Rodas:2023gma} and Roy-equation~\cite{Cao:2023ntr,Rodas:2024qhn} analyses of the lattice data.
}
\end{figure}

\begin{figure}
\includegraphics[width=0.3\textwidth]{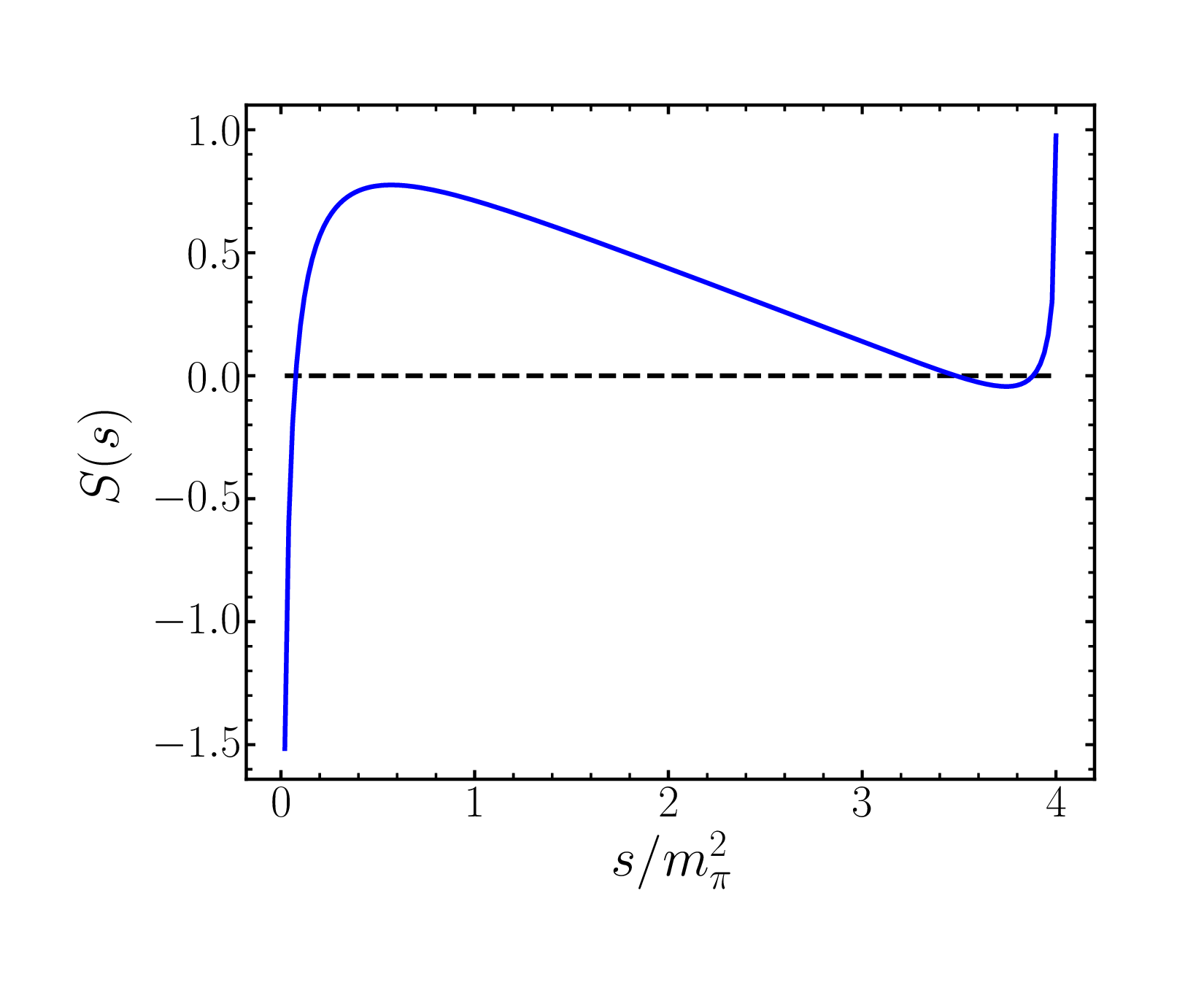}
\includegraphics[width=0.3\textwidth]{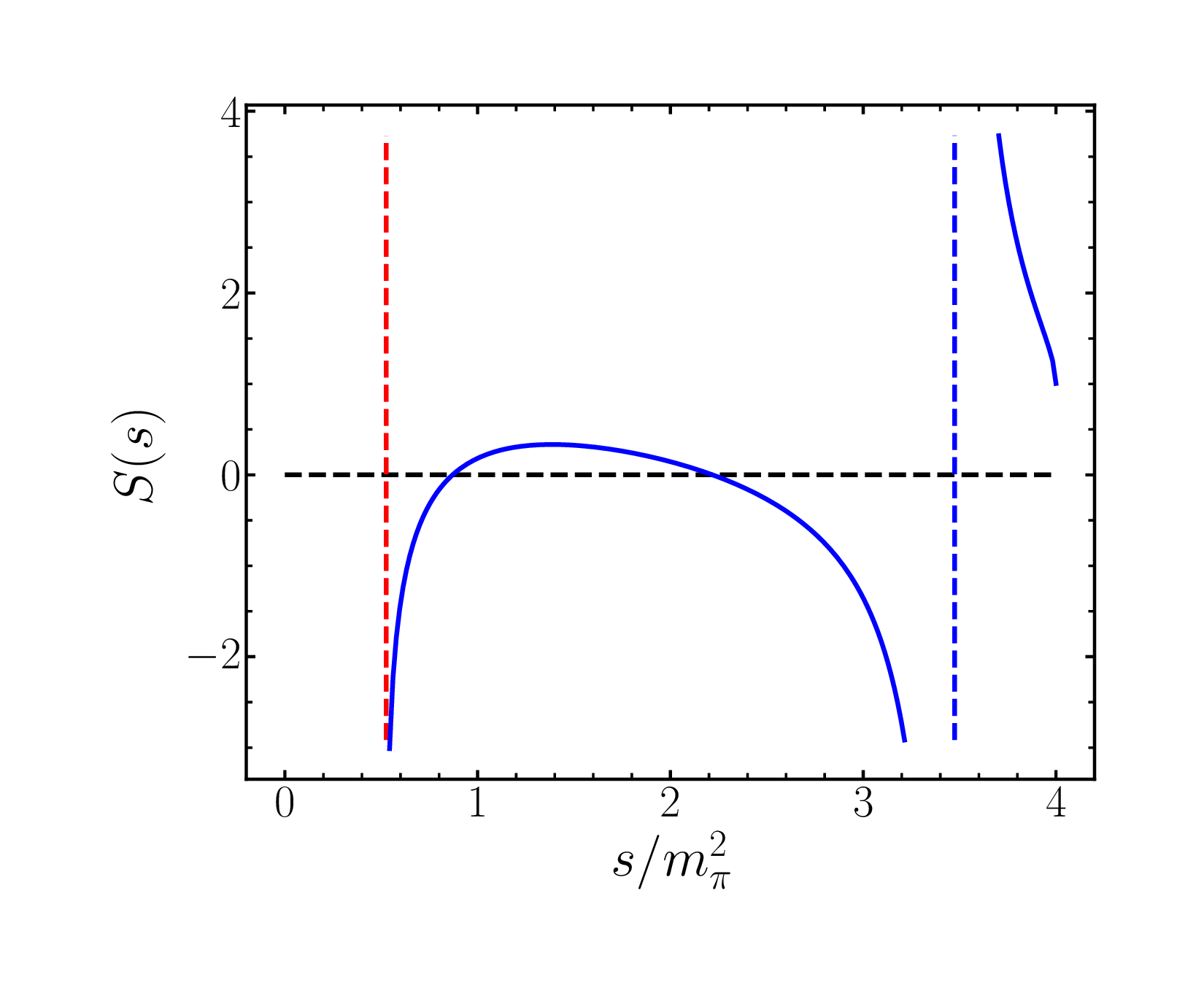}
\includegraphics[width=0.3\textwidth]{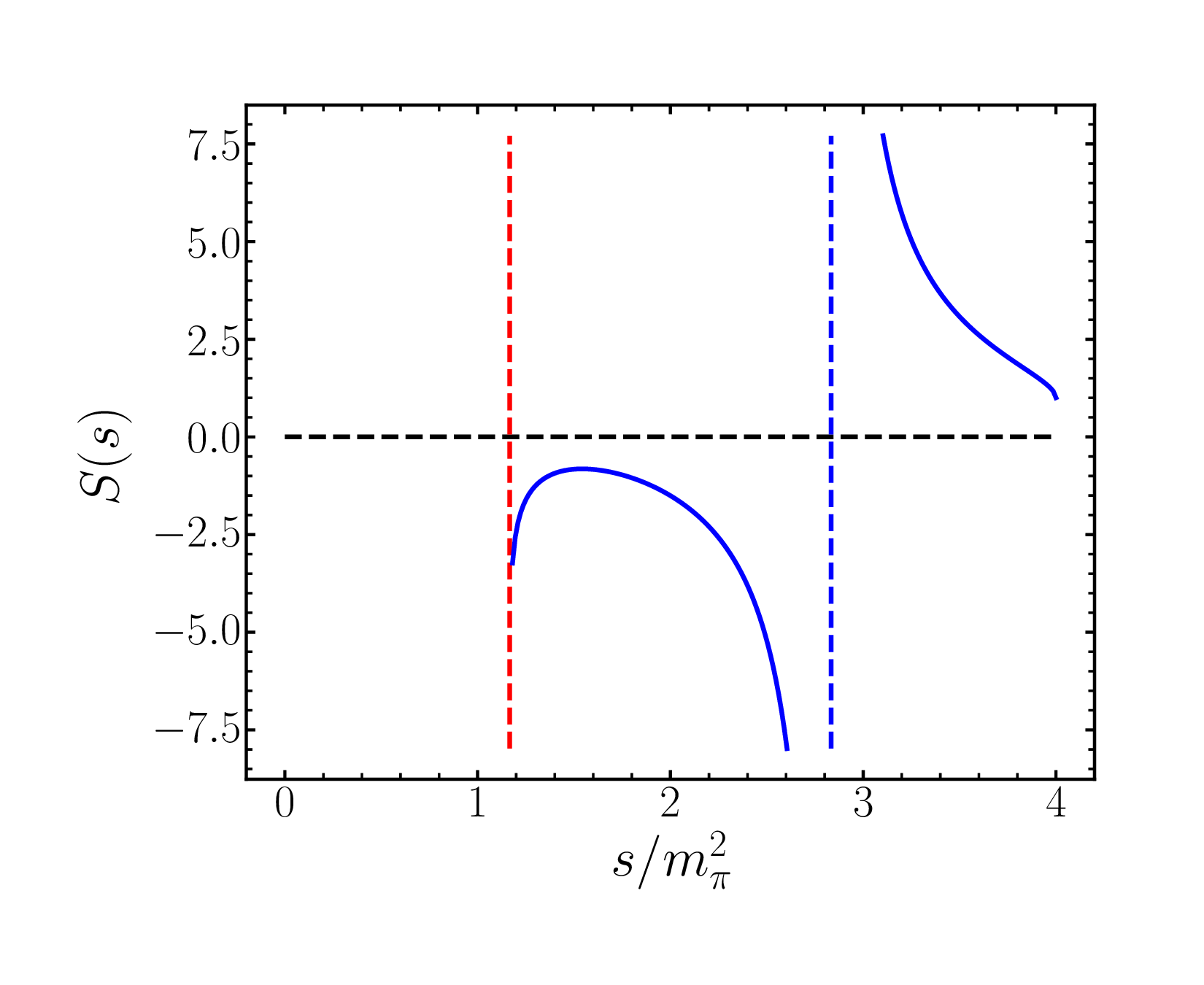}
\caption{ Blue solid line: physical-sheet S-matrix below threshold obtained in the
$N/D$ modified $O(N)$ model. Blue dashed line: position of the $\sigma$ bound state. Red dashed line: branch point of the left-hand cut ($s_L = 4m_\pi^2 - m_\sigma^2$). The left graph ($m_\pi =  207$ MeV) : two virtual states in the near threshold region and one additional virtual state pole generated close to the left-hand cut. The middle graph ($m_\pi =  283$ MeV) : one bound state with two virtual states. The right graph  ($m_\pi =391$ MeV):  two virtual state poles have become a pair of resonance poles.
\label{fig:S-below-th}
}
\end{figure}

The numerical results of the pole structure as $m_\pi$ increases are
shown in Fig.~\ref{fig:poletraj-ND}. In order to obtain a $\sigma$ pole close to the one in the leading order $O(N)$ amplitude at physical pion mass, the intrinsic scale $M$ is set to $1.5\,\mathrm{GeV}$ with $m_c \simeq 214 \,\mathrm{MeV}$ accordingly. 
When pion mass increases from the physical value $m_\pi = 139$ MeV, the
 $\sigma$ poles move towards the real axis and
at some point they hit the real axis and separate into two virtual state poles, with  one  moving
up (VS I) and the other moving down (VS II) along the real axis. The upper virtual state, VS I, moves across the
threshold to the physical Riemann sheet and then moves down along the real axis, 
hence becoming a bound state. At the same time, there will be a left-hand
cut generated with the branch point located at $s_L=4m_\pi^2-m_\sigma^2>0$ in the partial
wave amplitude, which comes from the $\sigma$-exchange in the $t$- and $u$-channels from crossing
symmetry.
During this course but before VS I turns into a bound state, an additional virtual state pole (VS III) is generated from the
left-hand cut, which was also implied in Ref.~\cite{Rodas:2024qhn}. With $m_\pi$ growing larger, VS II and VS III hit
each other, move into the complex plane, and become a pair of
subthreshold  poles. 
In fact,
in Ref.~\cite{Cao:2023ntr}, it was  shown that  besides a $\sigma$ bound
state below the threshold, there is also a pair of conjugate poles below the threshold when $m_\pi=391$ MeV, which corresponds to the pair of subthreshold poles in Fig.~\ref{fig:poletraj-ND}.
It was argued  that such subthreshold poles are inevitable due to the behavior of the left-hand cut
generated from the $\sigma$ exchange in crossed $t$- and $u$-channels. A more careful argument goes as follows. Considering  that the $S$-matrix near $s_L$ is mainly contributed by the
singular behavior of the branch point of the $\log$ cut generated from
the cross-channel $\sigma$ bound state pole, if the residue of the $\sigma$ pole is 
positive\footnote{ In
the nonrelativistic scattering theory, the residual of the scattering amplitude
is related to $-\gamma^2$ for $s$ wave, where $\gamma\in \mathbb R$ is defined
using the asymptotic behavior of the wave function
$\eta(r)\xlongrightarrow{r\to\infty} \gamma e^{-\alpha
r}$~\protect\cite{Taylor:1972pty}, thus the residue for the $S$-matrix
at the $\sigma$ bound state is positive. In relativistic quantum field theory, the positivity of the residue follows from unitarity.
},  which is always the case in
practice, the sign of the
partial wave $S$-matrix immediately above $s_L$ on the real axis
can be proved to be negative (opposite to the sign  of the residue). 
If $m_\pi$ is not large enough such
that VS II is still on the real axis, the $S$-matrix
below threshold between $s_L$ and the
$\sigma$ pole should have two zero points as shown in the middle
graph of Fig.~\ref{fig:S-below-th}. 
The two zeros of $S$-matrix mean
that besides VS II generated from the original $\sigma$
poles another virtual state pole VS III is generated from the branch point of
the left-hand cut\footnote{The second  sheet $S$-matrix is the inverse of the one defined on the physical sheet, owing to unitarity and analyticity.}.  As the pion mass increases further, the $S$-matrix
between the branch point and the bound state becomes smaller such that
VS II and VS III move towards each other, and then they hit
each other at some point  and move into the complex plane, becoming
a new resonance.  The final situation is illustrated in the right graph of Fig.~\ref{fig:S-below-th} and is what found at
$m_\pi=391\,\mathrm{MeV}$. Admittedly, both the above argument and the one in Ref.~\cite{Cao:2023ntr} can only ensure the existence of VS III when there is a $\sigma$ bound state.

However as pointed out above, the situation is a little different in $N/D$ modified $O(N)$ model, where the additional virtual state pole, VS III, is generated before the $\sigma$ becomes a bound state (see the left graph of Fig.~\ref{fig:S-below-th}), i.e. without the left-hand cut generated by the $\sigma$-bound-state exchange in the crossed-channels. The origin of this VS III state is found to be related with the interplay of Adler zero and the left-hand cut. 
As $m_\pi$ increases, we found that the real Adler zero, moving towards the left-hand cut, will hit the branch point $s_L = 0$ and go into the complex plane, becoming a pair of conjugate zeros. Remarkably, complex Adler zeros at $m_\pi \sim 391$ MeV were first found in Ref.~\cite{Cao:2023ntr} and real Adler zero can not be found in Ref.~\cite{Rodas:2024qhn} for $m_\pi \sim 283$ MeV, which may indicate that the above behavior of the Adler zero could be general.  
Then, it is straightforward to prove that VS III is generated when Adler zero passes $s_L$ as follows. The partial wave $S$-matrix is defined as $S(s) = 1 + 2i\rho(s) \mathcal T(s)$, thus $S(s) \to -\mathrm{sign}( \mathcal T(s)) \infty$ when $s \to s_L + 0^+$, as long as $\lim_{s \to s_L + 0^+} \mathcal T(s) \neq  0$.
\footnote{The situation here is similar to but much more complicated than that of physical $m_\pi$~\cite{Zhou:2004ms}.}
Then the Adler zero, i.e. the simple zero point of $\mathcal T(s)$, passing $s_L$, causes a sign change of $\mathcal T(s)$ from negative to positive and hence also a sign change of $S(s)$, at $s \to s_L + 0^+$.  This phenomenon has to occur before the $\sigma$ turns into a bound state, since according to the argument of the previous paragraph, the sign of $S$-matrix close to the branch point can not be flipped when $\sigma$ remains a bound state. Considering that the $S$-matrix equals to $1$ both at the threshold and at the Adler zero, there can only be even number of $S$-matrix zeros between these two points. Thus before the Adler zero passes $s_L$, if there are two virtual states, i.e. VS I and II, generated after the $\sigma$ resonance poles hit the real axis,   they must both be within that region. For smaller $m_\pi$ values, if there is no VS III and the Adler zero has not hit the left-hand cut branch point yet, thus $S(s)$ is zero-free between the Adler zero and $s_L$. Then $S(s)$ has to be positive on that interval. When Adler zero moves through the branch point, it causes the sign of  $S$-matrix near the left-hand cut changing from positive to negative.
This sign flip of $S(s)$ in the vicinity of $s_L$ indicates the generation of a $S$-matrix zero point from the branch point of the left-hand cut as shown in the left graph in Fig.~\ref{fig:S-below-th}, which exactly results in the appearance of VS III. 
As the pion mass grows further, the situation is similar to the description in the previous paragraph where VS I moves to the first sheet becoming a bound state, and VS II hit VS III  then both moving into the complex plane.  Thus, the existence of VS III and hence the subthreshold poles at large pion mass, e.g. $m_\pi \sim 391$ MeV, is actually the combined result of unitarity, crossing symmetry, Adler zero (which is a significant feature for low energy chiral dynamics) and the analyticity of
the $S$-matrix in this region.   

The $N/D$ modified $O(N)$ model reproduces the
picture derived from  Roy-equation analyses, which
demonstrates that the $\sigma$ particle in the $O(N)$ model really can
represent, \emph{at the qualitative level}, the $f_0$ state extracted by Roy equation from the
lattice data. Conversely, this also means that the lowest $f_0$ state
in the low energy $\pi\pi$ scattering really plays the similar role of the
 $\sigma$ in  $O(N)$ linear $\sigma$ model --- providing the vacuum
expectation value for spontaneous breaking of chiral symmetry.

\section {Sigma pole trajectory with temperature for different $m_{\pi}$ \label{sect:temperature}}
\begin{figure}[!htbp]       
    \centering
    \includegraphics[width=0.3\textwidth]{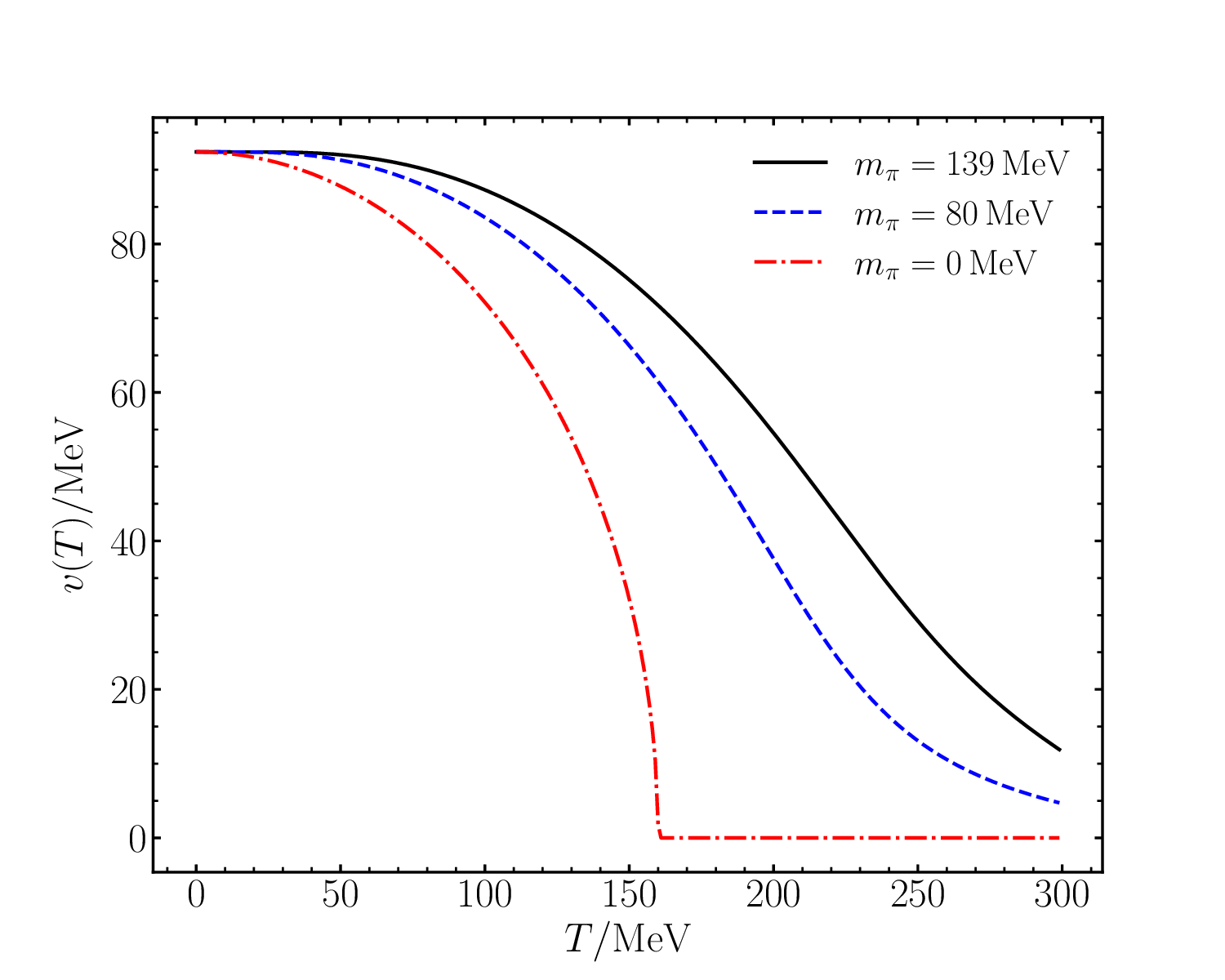}
    \includegraphics[width=0.3\textwidth]{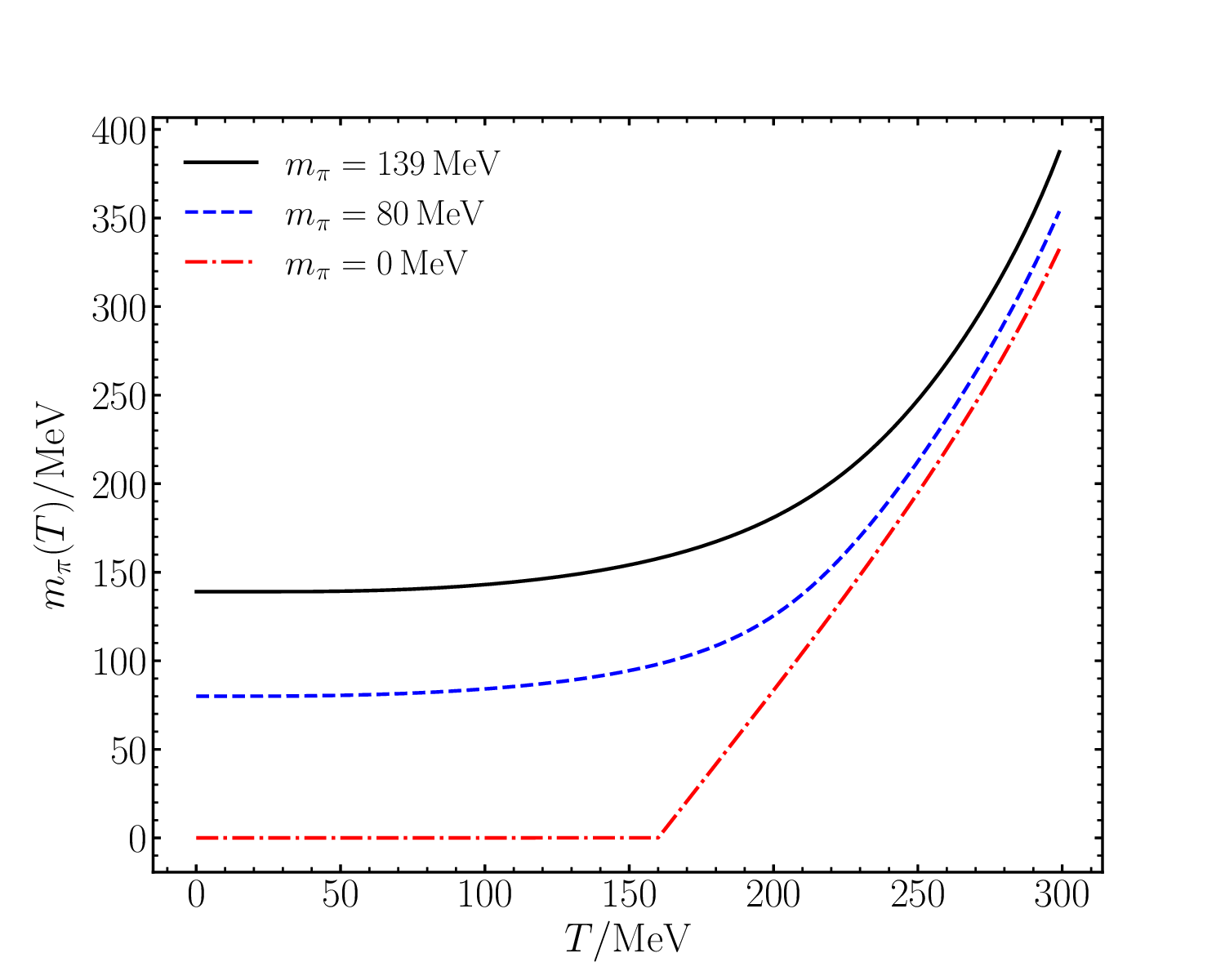}
    \caption{$v(T)$ (left) and $m_\pi(T)$ (right). The three cases are identified by $m_\pi(0)=139$, $80$ and $0$ MeV. In the chiral limit, there is a second-order phase transition with the critical temperature $T_c \simeq 160 \,\mathrm{MeV}$. With a nonzero $m_\pi(0)$, there is no explicit transition point.} 
    \label{fig: v_T and mpi_T}  
\end{figure}

It is also instructive to look at the $\sigma$ pole trajectory in a
finite temperature environment. 
It is well known that in the chiral
limit, i.e., when the pion is massless at zero-temperature, under high
temperature the system goes through a phase transition from the
chiral symmetry broken phase to the chiral symmetric phase
at a critical temperature $T_c$~\cite{Pisarski:1983ms,Bazavov:2011nk,HotQCD:2018pds,HotQCD:2019xnw} (e.g. for $O(N)$ model without explicit  symmetry breaking~\cite{Bochkarev:1995gi,Andersen:2004ae}, $T_c = \sqrt{12/N} f_\pi \simeq 160\,\mathrm{MeV}$, which can be easily read out in Fig.~\ref{fig: v_T and mpi_T}).
One would expect that above the phase transition temperature the massless pion gets massive and the sigma
particle would be degenerate with pions.  With explicit chiral
symmetry breaking where the pion has a mass at zero-temperature, there
is no explicit  phase transition point. But with temperature going
higher and higher, the system asymptotically approaches the chiral
symmetric phase, where the mass of the sigma tends to the pion mass. 
Notice that this whole picture can not be explicitly
realized in $\chi$PT, since it is constructed intrinsically
in the broken phase. This means that $\chi$PT is valid only for an
energy or temperature far below the chiral symmetry breaking
scale, which is determined by the VEV of the scalar field or $\bar q
q$. Under each of the circumstances: (a) the VEV approaches zero; (b) the energy scale or (c) the temperature,
becomes comparable with or goes beyond the chiral symmetry breaking scale, the whole
theory breaks down. For example, in Refs.~\cite{Gao:2019idb,Cortes:2015emo}, 
the sigma pole was found to be still in the complex plane of the second Riemann
sheet around $T=200$ MeV, which seems to be hopeless to restore the $O(4)$ 
symmetry in $\chi$PT around $T_c$.

The above picture can be easily seen in  $O(N)$ linear $\sigma$ model. 
The leading order effective potential at finite temperature $T$ can be obtained with imaginary-time formalism~\cite{Meyers-Ortmanns:1993dhx,Meyer-Ortmanns:1996ioo,Bochkarev:1995gi,Andersen:2004ae} by Wick rotation to Euclidean space and substituting the momentum integral with a sum over Matsubara frequencies $\omega_n =2\pi n T$, i.e.~$\int \frac{\mathrm d^4 k}{(2\pi)^4} f(k_0,\mathbf k) \to i T \sum_n \int \frac{\mathrm d^3 \mathbf k}{(2\pi)^3} f(k_0 = i \omega_n,\mathbf k)$ (see, e.g. Ref.~\cite{Bellac:2011kqa} for details). 
The renormalization condition is chosen the same as the $T=0$ case. Then by
minimizing the effective potential, we can obtain the gap
equations for $v$ and $m_\pi^2$ as functions of
temperature~\cite{Andersen:2004ae}:
\begin{align}\label{eq:finiteT Gap eq 1}
    v^2(T) &= f_\pi^2 +\frac{N}{16\pi^2}\left( m_\pi^2 \log\frac{m_\pi^2}{M^2} - m^2_\pi(T)  \log\frac{m_\pi^2(T) }{M^2}  \right) - N A^{T\neq 0} \left(m_\pi^2(T)\right)\,, \\
    \alpha &= v(T)m_\pi^2(T) \,,
    \label{eq:finiteT Gap eq 2}
\end{align}
where $v(0)=f_\pi$ and $m_\pi(0)=m_\pi$ are set to zero-temperature values. The function $A^{T\neq 0}$ is defined as the finite temperature contribution to the tadpole integral encountered in the derivation of Eq.~\eqref{eq:zeroT Gap eq 1},
\begin{align}
    A^{T\neq 0} \left(m_\pi^2(T)\right) = \int^\infty_0 \frac{\mathrm d k}{2\pi^2} \frac{k^2 n_B(\omega_k)}{\omega_k}\,,
\end{align}
where $\beta =1/T$, $\omega_k = \sqrt{k^2+m_\pi^2(T)}$ and $n_B(\omega_k) = (e^{\beta\omega_k}-1)^{-1}$ is the Bose-Einstein distribution. With different values of $m_\pi(0)$, denoting the magnitude of explicit breaking, the solutions of $v(T)$ and $m_\pi(T)$ are shown in Fig.~\ref{fig: v_T and mpi_T}.

\begin{figure}[!htbp]       
    \centering
    \includegraphics[width=0.3\textwidth]{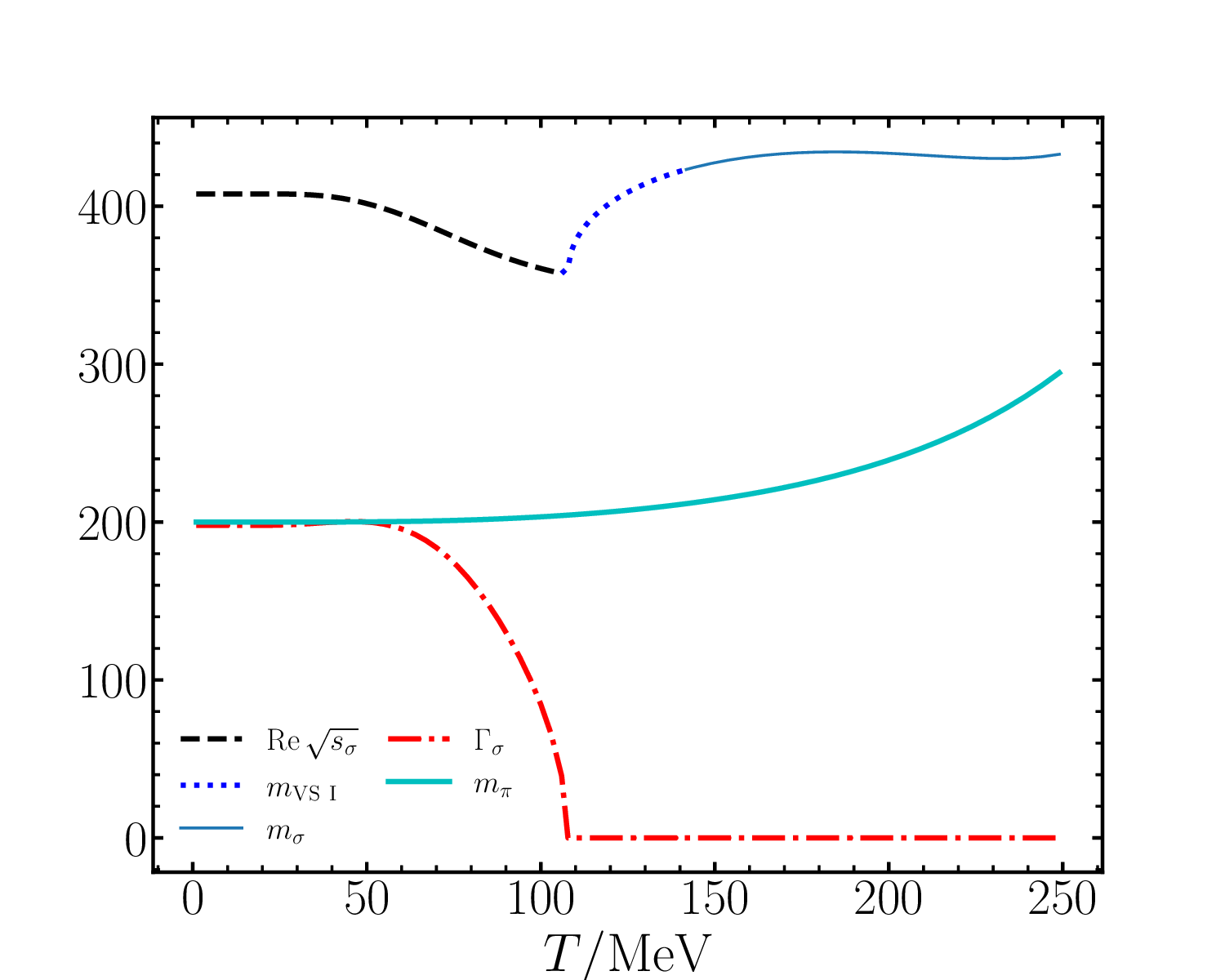}
    \includegraphics[width=0.3\textwidth]{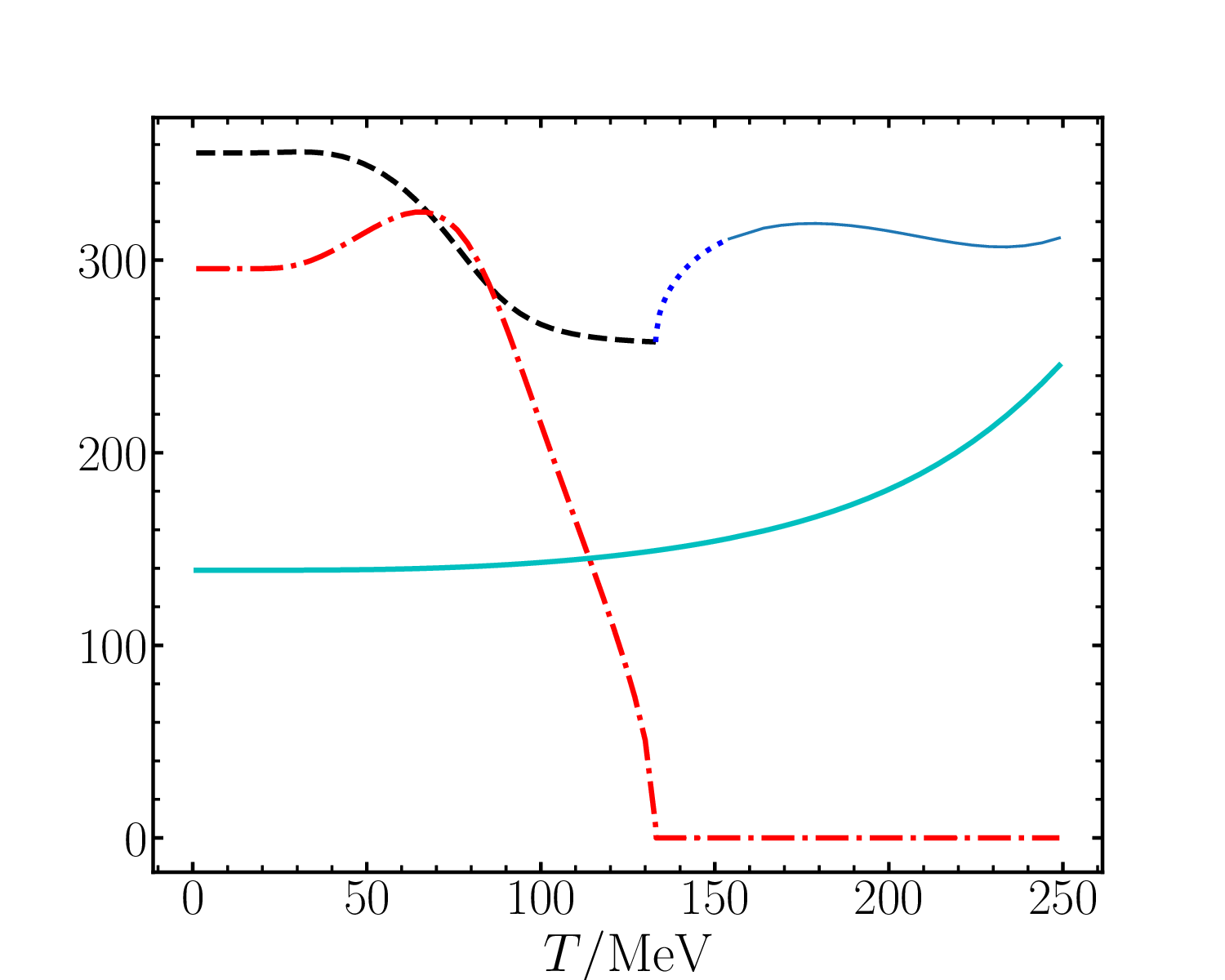}
    \includegraphics[width=0.3\textwidth]{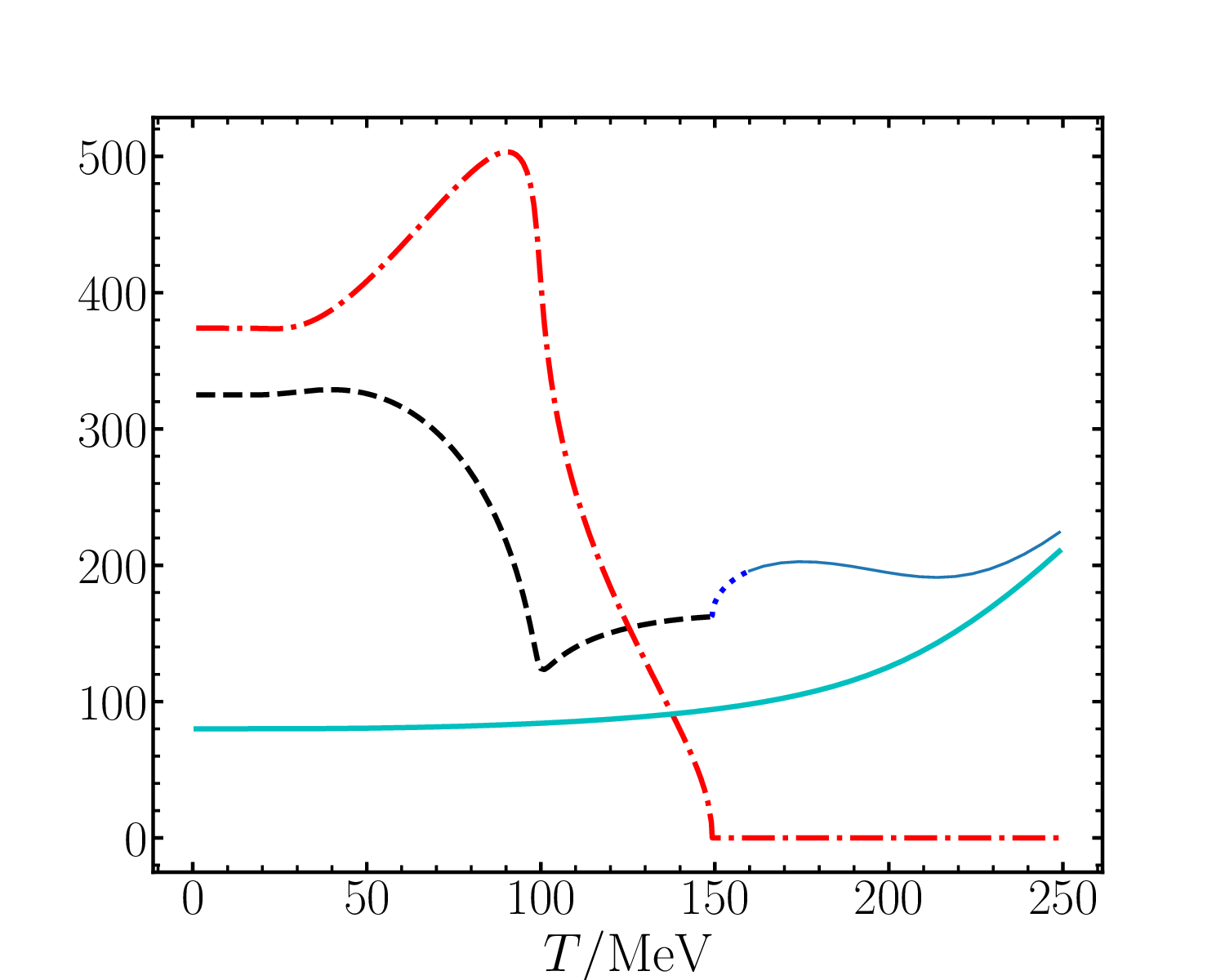}
    \caption{ The mass and width of $\sigma$ pole with varying
temperature compared with $m_\pi(T)$. From left to right, 
$m_\pi(0)=200$, $139$ and $80$ MeV respectively. The lower virtual state pole VS II is not shown above.} 
    \label{fig: sigma mass and width finiteT}  
\end{figure}

To study the spectrum at finite temperature in the center of
mass (CM) frame, the scattering amplitude is generalized to be
the amputated four-point Green's function with the external momenta  analytically continued back to on-shell momenta in Minkowski space after the Matsubara sum~\cite{Quack:1994vc,Kaiser:1999mt,GomezNicola:2002tn,Gao:2019idb}.
The leading $1/N$ order $\pi\pi$ scattering amplitude with finite temperature can be expressed as,
\begin{align}\label{eq:Tamp fniteT}
     \mathcal T^{T}_{00}(s) =  -\frac{1}{32\pi} \frac{s -m_\pi^2(T) }{ \left(s -m_\pi^2(T)\right)\,
     B^{T}\left(s,m_\pi(T),M\right) -v^2(T)/N}\,,
\end{align}
where $B^{T}\left(s,m_\pi(T),M\right)$, the finite temperature version
of $B\left(s,m_\pi,M\right)$ defined in Eq.~\eqref{eq: def of ren-Bfun},
can be obtained by standard calculations,
\begin{align}
    B^{T}\left(s,m_\pi(T),M\right) &= B\left(s,m_\pi(T),M\right) + B^{T\neq 0}\left(s,m_\pi(T)\right)\,, \\
    B^{T\neq 0}\left(s,m_\pi(T)\right) &= \int^{\infty}_0 \frac{\mathrm d k \, k^2}{8\pi^2 \omega_k^2}
    n_B(\omega_k)\left(  \frac{1}{E+2\omega_k} - \frac{1}{E-2\omega_k} \right) \,,
\end{align}
with $B^{T\neq 0}$ evaluated in the CM frame and $s = E^2$.
The $\sigma$ resonance pole can be obtained from the zero point of the
denominator of $\mathcal T^{T}_{00}$ on the second Riemann sheet. The mass
and width for the $\sigma$ pole with varying temperature and with
$m_\pi(0)=200$, $139$ and $80$ MeV respectively are illustrated in
Fig.~\ref{fig: sigma mass and width finiteT}, in which $m_\sigma$ and $\Gamma_\sigma$ are also compared to
the behavior of $m_\pi$ at finite temperature for each case. As temperature
increases, $\sigma$ resonance firstly becomes even broader due to the
growth in phase space caused by the Bose-Einstein distribution
$n_B(\omega_k)$, and then $\Gamma_\sigma$ drops rapidly to zero
when $\sigma$ turns into a pair of virtual state poles. For the present purpose, to demonstrate the asymptotic degeneration of  $\sigma$ and $\pi$'s, we only keep track of the virtual state pole, VS I (named in the same way as the zero-temperature case), which moves up through threshold to the physical sheet and becomes a $\sigma$ bound state at around $T_c$.  
Then $m_\sigma$ changes gradually and moves closer to
$m_\pi(T)$ when $T>T_c$. Furthermore, $m_\sigma$ and $m_\pi(T)$
asymptotically tends to converge at $T\gg T_c$, which is as expected by 
the restoration of  chiral symmetry at high temperature limit. 

\section {Conclusions and discussions \label{sect:conclusion}}
In this paper, we have investigated the $O(N)$ linear $\sigma$ model with
varying $m_\pi$ to  reproduce the subthreshold pole structure of the $IJ=00$ channel
$\pi\pi$ scattering amplitude, recently found using the Roy equation in
analyzing the lattice phase shifts at several unphysical $m_\pi$ values. By using the
$N/D$ method to partially recover  crossing symmetry of the partial
wave amplitude in $O(N)$ model, the Roy-equation analysis results
can be roughly reproduced.   The pole trajectory illustrates the picture
that with large $m_\pi$, as the Adler zero goes into the complex plane before $\sigma$ becomes a bound state,
besides the virtual state accompanying the $\sigma$ bound state,
another virtual state pole will originate from the left-hand cut. As
$m_\pi$ grows larger, the two virtual state poles hit each other and
scattered into the complex plane and become a pair of resonance poles. 

The consistency of $O(N)$ model with the Roy-equation analyses
reveals that the lowest $f_0$ state from the Roy-equation analyses of the
lattice data can be fairly described by the $\sigma$ field in 
$O(N)$ linear $\sigma$ model, thus plays the same role in the chiral
symmetry breaking as the $\sigma$ particle in $O(N)$ model.

The $\sigma$ pole behavior with varying temperature is also discussed
in the $O(N)$ model with different pion masses. It is shown that with
higher temperature the sigma resonance pole also moves to the real axis and 
becomes a pair of virtual states. Then the upper one moves across 
threshold to the first Riemann sheet and turns into a bound state. 
With much higher temperatures, the mass of the sigma and pion would
tend to come closer and closer, which is as expected by chiral
symmetry restoration in the high temperature limit.

It is worth mentioning that $O(N)$ model itself suffers from the
vacuum stability problem. 
In short, the original $O(N)$ model effective potential is obtained by
solving the auxiliary field $\chi(\phi^2)$  as a function of $\phi_a$
from one of the gap equations $\frac{\partial V}{\partial \chi} =0$,
which has two branches of solutions: one with ordinary chiral symmetry
breaking vacuum (i.e. $m_\pi\to 0$ when $\alpha \to 0$),
the other  with a larger pion
mass which restores chiral symmetry ($v=0$ but $m_\pi\neq 0$) in the
$\alpha\to 0$ limit~\cite{Coleman:1974jh,Kobayashi:1975ev,Abbott:1975bn,Linde:1976qh,Bardeen:1983st,Bardeen:1986td}.
In perturbation theory, $V(\phi)$ is expanded around the vacuum which
is a stationary point of the effective potential. Whether it is a
local minimum, local maximum or saddle point can be determined by the Hessian matrix of $V(\phi)$. 
After a careful investigation we
found that the vacuum chosen at $v=f_\pi$ with $\chi(v^2)=m_\pi^2$ remains
a local minimum on the first branch when $m_\pi<M/\sqrt e$. However for larger
pion mass, the stable minimum exists only on the second branch.  This
is a warning that when $m_\pi$ gets large, the vacuum of the $O(N)$
model becomes unstable -- a phenomenon not known in our knowledge of
QCD\footnote{The situation of $\chi$PT is even worse due to the Ostr\protect{\"o}gradski instability, such that the $\chi$PT Hamiltonian is unbounded from below.}.
A similar problem also happens at high temperature: when
$T\gg T_c$, the effective potential no longer provides a local
minimum on its first branch. Instead, the local minimum will move to
the second branch and become a saddle point. In spite of these
deficiencies, we insist on the opinion that the linear $\sigma$ model
on the first branch provides the correct picture for describing low
energy QCD, since high dimensional terms and other resonance terms,
missed in the present discussion based on the toy linear sigma model,
may alleviate the vacuum stability problem. After all, it is worthwhile to look at
the $m_\pi$ dependence of the effective potential which is not
discussed in the literature yet, to the best of our knowledge.  We will present the details of these discussions elsewhere.

Another direction to be explored is to look at the behavior of the
$N^*(920)$ recently found in the $\pi N$ scattering from the Roy-equation analysis~\cite{Cao:2022zhn,Wang:2017agd,Hoferichter:2023mgy}
under different pion masses or temperatures.
The role $N^*(920)$ plays in the $\pi N$ scattering  is in some sense
similar to the $\sigma$ particle in $\pi\pi$ scattering. If it shares
similar properties at higher temperatures, one
would expect that $N^*(920)$ would become a bound state and the parity partner of the nucleon,
and thus may play an important role in physics.

\begin{acknowledgments}
 This work is supported by China National Natural Science Foundation
under contract No. 12335002, 12375078, 11975028.
This work is also supported by “the Fundamental Research Funds for the Central Universities”.
\end{acknowledgments}

\bibliographystyle{apsrev4-2}
\bibliography{ref-ON}

\begin{thebibliography}{84}%
\makeatletter
\providecommand \@ifxundefined [1]{%
 \@ifx{#1\undefined}
}%
\providecommand \@ifnum [1]{%
 \ifnum #1\expandafter \@firstoftwo
 \else \expandafter \@secondoftwo
 \fi
}%
\providecommand \@ifx [1]{%
 \ifx #1\expandafter \@firstoftwo
 \else \expandafter \@secondoftwo
 \fi
}%
\providecommand \natexlab [1]{#1}%
\providecommand \enquote  [1]{``#1''}%
\providecommand \bibnamefont  [1]{#1}%
\providecommand \bibfnamefont [1]{#1}%
\providecommand \citenamefont [1]{#1}%
\providecommand \href@noop [0]{\@secondoftwo}%
\providecommand \href [0]{\begingroup \@sanitize@url \@href}%
\providecommand \@href[1]{\@@startlink{#1}\@@href}%
\providecommand \@@href[1]{\endgroup#1\@@endlink}%
\providecommand \@sanitize@url [0]{\catcode `\\12\catcode `\$12\catcode
  `\&12\catcode `\#12\catcode `\^12\catcode `\_12\catcode `\%12\relax}%
\providecommand \@@startlink[1]{}%
\providecommand \@@endlink[0]{}%
\providecommand \url  [0]{\begingroup\@sanitize@url \@url }%
\providecommand \@url [1]{\endgroup\@href {#1}{\urlprefix }}%
\providecommand \urlprefix  [0]{URL }%
\providecommand \Eprint [0]{\href }%
\providecommand \doibase [0]{https://doi.org/}%
\providecommand \selectlanguage [0]{\@gobble}%
\providecommand \bibinfo  [0]{\@secondoftwo}%
\providecommand \bibfield  [0]{\@secondoftwo}%
\providecommand \translation [1]{[#1]}%
\providecommand \BibitemOpen [0]{}%
\providecommand \bibitemStop [0]{}%
\providecommand \bibitemNoStop [0]{.\EOS\space}%
\providecommand \EOS [0]{\spacefactor3000\relax}%
\providecommand \BibitemShut  [1]{\csname bibitem#1\endcsname}%
\let\auto@bib@innerbib\@empty
%</preamble>
\bibitem [{\citenamefont {Gell-Mann}\ and\ \citenamefont
  {L\'evy}(1960)}]{Gell-Mann:1960mvl}%
  \BibitemOpen
  \bibfield  {author} {\bibinfo {author} {\bibfnamefont {M.}~\bibnamefont
  {Gell-Mann}}\ and\ \bibinfo {author} {\bibfnamefont {M.}~\bibnamefont
  {L\'evy}},\ }\href {https://doi.org/10.1007/BF02859738} {\bibfield  {journal}
  {\bibinfo  {journal} {Nuovo Cim.}\ }\textbf {\bibinfo {volume} {16}},\
  \bibinfo {pages} {705} (\bibinfo {year} {1960})}\BibitemShut {NoStop}%
\bibitem [{\citenamefont {Islam}\ and\ \citenamefont
  {Pinon}(1964)}]{Islam:1964zz}%
  \BibitemOpen
  \bibfield  {author} {\bibinfo {author} {\bibfnamefont {M.~M.}\ \bibnamefont
  {Islam}}\ and\ \bibinfo {author} {\bibfnamefont {R.}~\bibnamefont {Pinon}},\
  }\href {https://doi.org/10.1103/PhysRevLett.12.310} {\bibfield  {journal}
  {\bibinfo  {journal} {Phys. Rev. Lett.}\ }\textbf {\bibinfo {volume} {12}},\
  \bibinfo {pages} {310} (\bibinfo {year} {1964})}\BibitemShut {NoStop}%
\bibitem [{\citenamefont {Patil}(1964)}]{Patil:1964zz}%
  \BibitemOpen
  \bibfield  {author} {\bibinfo {author} {\bibfnamefont {S.~H.}\ \bibnamefont
  {Patil}},\ }\href {https://doi.org/10.1103/PhysRevLett.13.261} {\bibfield
  {journal} {\bibinfo  {journal} {Phys. Rev. Lett.}\ }\textbf {\bibinfo
  {volume} {13}},\ \bibinfo {pages} {261} (\bibinfo {year} {1964})}\BibitemShut
  {NoStop}%
\bibitem [{\citenamefont {Hagopian}\ \emph {et~al.}(1966)\citenamefont
  {Hagopian}, \citenamefont {Selove}, \citenamefont {Alitti},\ and\
  \citenamefont {Baton}}]{Hagopian:1966zz}%
  \BibitemOpen
  \bibfield  {author} {\bibinfo {author} {\bibfnamefont {V.}~\bibnamefont
  {Hagopian}}, \bibinfo {author} {\bibfnamefont {W.}~\bibnamefont {Selove}},
  \bibinfo {author} {\bibfnamefont {J.}~\bibnamefont {Alitti}},\ and\ \bibinfo
  {author} {\bibfnamefont {J.~P.}\ \bibnamefont {Baton}},\ }\href
  {https://doi.org/10.1103/PhysRev.145.1128} {\bibfield  {journal} {\bibinfo
  {journal} {Phys. Rev.}\ }\textbf {\bibinfo {volume} {145}},\ \bibinfo {pages}
  {1128} (\bibinfo {year} {1966})}\BibitemShut {NoStop}%
\bibitem [{\citenamefont {Coleman}\ \emph {et~al.}(1969)\citenamefont
  {Coleman}, \citenamefont {Wess},\ and\ \citenamefont
  {Zumino}}]{Coleman:1969sm}%
  \BibitemOpen
  \bibfield  {author} {\bibinfo {author} {\bibfnamefont {S.~R.}\ \bibnamefont
  {Coleman}}, \bibinfo {author} {\bibfnamefont {J.}~\bibnamefont {Wess}},\ and\
  \bibinfo {author} {\bibfnamefont {B.}~\bibnamefont {Zumino}},\ }\href
  {https://doi.org/10.1103/PhysRev.177.2239} {\bibfield  {journal} {\bibinfo
  {journal} {Phys. Rev.}\ }\textbf {\bibinfo {volume} {177}},\ \bibinfo {pages}
  {2239} (\bibinfo {year} {1969})}\BibitemShut {NoStop}%
\bibitem [{\citenamefont {Callan}\ \emph {et~al.}(1969)\citenamefont {Callan},
  \citenamefont {Coleman}, \citenamefont {Wess},\ and\ \citenamefont
  {Zumino}}]{Callan:1969sn}%
  \BibitemOpen
  \bibfield  {author} {\bibinfo {author} {\bibfnamefont {C.~G.}\ \bibnamefont
  {Callan}, \bibfnamefont {Jr.}}, \bibinfo {author} {\bibfnamefont {S.~R.}\
  \bibnamefont {Coleman}}, \bibinfo {author} {\bibfnamefont {J.}~\bibnamefont
  {Wess}},\ and\ \bibinfo {author} {\bibfnamefont {B.}~\bibnamefont {Zumino}},\
  }\href {https://doi.org/10.1103/PhysRev.177.2247} {\bibfield  {journal}
  {\bibinfo  {journal} {Phys. Rev.}\ }\textbf {\bibinfo {volume} {177}},\
  \bibinfo {pages} {2247} (\bibinfo {year} {1969})}\BibitemShut {NoStop}%
\bibitem [{\citenamefont {Gasser}\ and\ \citenamefont
  {Leutwyler}(1984)}]{Gasser:1983yg}%
  \BibitemOpen
  \bibfield  {author} {\bibinfo {author} {\bibfnamefont {J.}~\bibnamefont
  {Gasser}}\ and\ \bibinfo {author} {\bibfnamefont {H.}~\bibnamefont
  {Leutwyler}},\ }\href {https://doi.org/10.1016/0003-4916(84)90242-2}
  {\bibfield  {journal} {\bibinfo  {journal} {Annals Phys.}\ }\textbf {\bibinfo
  {volume} {158}},\ \bibinfo {pages} {142} (\bibinfo {year}
  {1984})}\BibitemShut {NoStop}%
\bibitem [{\citenamefont {Gasser}\ and\ \citenamefont
  {Leutwyler}(1985)}]{Gasser:1984gg}%
  \BibitemOpen
  \bibfield  {author} {\bibinfo {author} {\bibfnamefont {J.}~\bibnamefont
  {Gasser}}\ and\ \bibinfo {author} {\bibfnamefont {H.}~\bibnamefont
  {Leutwyler}},\ }\href {https://doi.org/10.1016/0550-3213(85)90492-4}
  {\bibfield  {journal} {\bibinfo  {journal} {Nucl. Phys. B}\ }\textbf
  {\bibinfo {volume} {250}},\ \bibinfo {pages} {465} (\bibinfo {year}
  {1985})}\BibitemShut {NoStop}%
\bibitem [{\citenamefont {Ecker}\ \emph {et~al.}(1989)\citenamefont {Ecker},
  \citenamefont {Gasser}, \citenamefont {Pich},\ and\ \citenamefont
  {de~Rafael}}]{Ecker:1988te}%
  \BibitemOpen
  \bibfield  {author} {\bibinfo {author} {\bibfnamefont {G.}~\bibnamefont
  {Ecker}}, \bibinfo {author} {\bibfnamefont {J.}~\bibnamefont {Gasser}},
  \bibinfo {author} {\bibfnamefont {A.}~\bibnamefont {Pich}},\ and\ \bibinfo
  {author} {\bibfnamefont {E.}~\bibnamefont {de~Rafael}},\ }\href
  {https://doi.org/10.1016/0550-3213(89)90346-5} {\bibfield  {journal}
  {\bibinfo  {journal} {Nucl. Phys. B}\ }\textbf {\bibinfo {volume} {321}},\
  \bibinfo {pages} {311} (\bibinfo {year} {1989})}\BibitemShut {NoStop}%
\bibitem [{\citenamefont {Donoghue}\ \emph {et~al.}(1989)\citenamefont
  {Donoghue}, \citenamefont {Ramirez},\ and\ \citenamefont
  {Valencia}}]{Donoghue:1988ed}%
  \BibitemOpen
  \bibfield  {author} {\bibinfo {author} {\bibfnamefont {J.~F.}\ \bibnamefont
  {Donoghue}}, \bibinfo {author} {\bibfnamefont {C.}~\bibnamefont {Ramirez}},\
  and\ \bibinfo {author} {\bibfnamefont {G.}~\bibnamefont {Valencia}},\ }\href
  {https://doi.org/10.1103/PhysRevD.39.1947} {\bibfield  {journal} {\bibinfo
  {journal} {Phys. Rev. D}\ }\textbf {\bibinfo {volume} {39}},\ \bibinfo
  {pages} {1947} (\bibinfo {year} {1989})}\BibitemShut {NoStop}%
\bibitem [{\citenamefont {Guo}\ \emph {et~al.}(2007{\natexlab{a}})\citenamefont
  {Guo}, \citenamefont {Sanz~Cillero},\ and\ \citenamefont
  {Zheng}}]{Guo:2007ff}%
  \BibitemOpen
  \bibfield  {author} {\bibinfo {author} {\bibfnamefont {Z.~H.}\ \bibnamefont
  {Guo}}, \bibinfo {author} {\bibfnamefont {J.~J.}\ \bibnamefont
  {Sanz~Cillero}},\ and\ \bibinfo {author} {\bibfnamefont {H.~Q.}\ \bibnamefont
  {Zheng}},\ }\href {https://doi.org/10.1088/1126-6708/2007/06/030} {\bibfield
  {journal} {\bibinfo  {journal} {JHEP}\ }\textbf {\bibinfo {volume} {06}},\
  \bibinfo {pages} {030}},\ \Eprint {https://arxiv.org/abs/hep-ph/0701232}
  {arXiv:hep-ph/0701232} \BibitemShut {NoStop}%
\bibitem [{\citenamefont {Guo}\ \emph {et~al.}(2008)\citenamefont {Guo},
  \citenamefont {Sanz-Cillero},\ and\ \citenamefont {Zheng}}]{Guo:2007hm}%
  \BibitemOpen
  \bibfield  {author} {\bibinfo {author} {\bibfnamefont {Z.~H.}\ \bibnamefont
  {Guo}}, \bibinfo {author} {\bibfnamefont {J.~J.}\ \bibnamefont
  {Sanz-Cillero}},\ and\ \bibinfo {author} {\bibfnamefont {H.~Q.}\ \bibnamefont
  {Zheng}},\ }\href {https://doi.org/10.1016/j.physletb.2008.01.073} {\bibfield
   {journal} {\bibinfo  {journal} {Phys. Lett. B}\ }\textbf {\bibinfo {volume}
  {661}},\ \bibinfo {pages} {342} (\bibinfo {year} {2008})},\ \Eprint
  {https://arxiv.org/abs/0710.2163} {arXiv:0710.2163 [hep-ph]} \BibitemShut
  {NoStop}%
\bibitem [{\citenamefont {Pelaez}(2004)}]{Pelaez:2004xp}%
  \BibitemOpen
  \bibfield  {author} {\bibinfo {author} {\bibfnamefont {J.~R.}\ \bibnamefont
  {Pelaez}},\ }\href {https://doi.org/10.1142/S0217732304016160} {\bibfield
  {journal} {\bibinfo  {journal} {Mod. Phys. Lett. A}\ }\textbf {\bibinfo
  {volume} {19}},\ \bibinfo {pages} {2879} (\bibinfo {year} {2004})},\ \Eprint
  {https://arxiv.org/abs/hep-ph/0411107} {arXiv:hep-ph/0411107} \BibitemShut
  {NoStop}%
\bibitem [{\citenamefont {Gomez~Nicola}\ and\ \citenamefont
  {Pelaez}(2002)}]{GomezNicola:2001as}%
  \BibitemOpen
  \bibfield  {author} {\bibinfo {author} {\bibfnamefont {A.}~\bibnamefont
  {Gomez~Nicola}}\ and\ \bibinfo {author} {\bibfnamefont {J.~R.}\ \bibnamefont
  {Pelaez}},\ }\href {https://doi.org/10.1103/PhysRevD.65.054009} {\bibfield
  {journal} {\bibinfo  {journal} {Phys. Rev. D}\ }\textbf {\bibinfo {volume}
  {65}},\ \bibinfo {pages} {054009} (\bibinfo {year} {2002})},\ \Eprint
  {https://arxiv.org/abs/hep-ph/0109056} {arXiv:hep-ph/0109056} \BibitemShut
  {NoStop}%
\bibitem [{\citenamefont {Qin}\ \emph {et~al.}(2002)\citenamefont {Qin},
  \citenamefont {Deng}, \citenamefont {Xiao},\ and\ \citenamefont
  {Zheng}}]{Qin:2002hk}%
  \BibitemOpen
  \bibfield  {author} {\bibinfo {author} {\bibfnamefont {G.-Y.}\ \bibnamefont
  {Qin}}, \bibinfo {author} {\bibfnamefont {W.~Z.}\ \bibnamefont {Deng}},
  \bibinfo {author} {\bibfnamefont {Z.}~\bibnamefont {Xiao}},\ and\ \bibinfo
  {author} {\bibfnamefont {H.~Q.}\ \bibnamefont {Zheng}},\ }\href
  {https://doi.org/10.1016/S0370-2693(02)02312-2} {\bibfield  {journal}
  {\bibinfo  {journal} {Phys. Lett. B}\ }\textbf {\bibinfo {volume} {542}},\
  \bibinfo {pages} {89} (\bibinfo {year} {2002})},\ \Eprint
  {https://arxiv.org/abs/hep-ph/0205214} {arXiv:hep-ph/0205214} \BibitemShut
  {NoStop}%
\bibitem [{\citenamefont {Yao}\ \emph {et~al.}(2021)\citenamefont {Yao},
  \citenamefont {Dai}, \citenamefont {Zheng},\ and\ \citenamefont
  {Zhou}}]{Yao:2020bxx}%
  \BibitemOpen
  \bibfield  {author} {\bibinfo {author} {\bibfnamefont {D.-L.}\ \bibnamefont
  {Yao}}, \bibinfo {author} {\bibfnamefont {L.-Y.}\ \bibnamefont {Dai}},
  \bibinfo {author} {\bibfnamefont {H.-Q.}\ \bibnamefont {Zheng}},\ and\
  \bibinfo {author} {\bibfnamefont {Z.-Y.}\ \bibnamefont {Zhou}},\ }\href
  {https://doi.org/10.1088/1361-6633/abfa6f} {\bibfield  {journal} {\bibinfo
  {journal} {Rept. Prog. Phys.}\ }\textbf {\bibinfo {volume} {84}},\ \bibinfo
  {pages} {076201} (\bibinfo {year} {2021})},\ \Eprint
  {https://arxiv.org/abs/2009.13495} {arXiv:2009.13495 [hep-ph]} \BibitemShut
  {NoStop}%
\bibitem [{\citenamefont {Xiao}\ and\ \citenamefont
  {Zheng}(2001)}]{Xiao:2000kx}%
  \BibitemOpen
  \bibfield  {author} {\bibinfo {author} {\bibfnamefont {Z.}~\bibnamefont
  {Xiao}}\ and\ \bibinfo {author} {\bibfnamefont {H.~Q.}\ \bibnamefont
  {Zheng}},\ }\href {https://doi.org/10.1016/S0375-9474(01)01100-9} {\bibfield
  {journal} {\bibinfo  {journal} {Nucl. Phys. A}\ }\textbf {\bibinfo {volume}
  {695}},\ \bibinfo {pages} {273} (\bibinfo {year} {2001})},\ \Eprint
  {https://arxiv.org/abs/hep-ph/0011260} {arXiv:hep-ph/0011260} \BibitemShut
  {NoStop}%
\bibitem [{\citenamefont {Zheng}\ \emph
  {et~al.}(2004{\natexlab{a}})\citenamefont {Zheng}, \citenamefont {Zhou},
  \citenamefont {Qin}, \citenamefont {Xiao}, \citenamefont {Wang},\ and\
  \citenamefont {Wu}}]{Zheng:2003rw}%
  \BibitemOpen
  \bibfield  {author} {\bibinfo {author} {\bibfnamefont {H.~Q.}\ \bibnamefont
  {Zheng}}, \bibinfo {author} {\bibfnamefont {Z.~Y.}\ \bibnamefont {Zhou}},
  \bibinfo {author} {\bibfnamefont {G.~Y.}\ \bibnamefont {Qin}}, \bibinfo
  {author} {\bibfnamefont {Z.}~\bibnamefont {Xiao}}, \bibinfo {author}
  {\bibfnamefont {J.~J.}\ \bibnamefont {Wang}},\ and\ \bibinfo {author}
  {\bibfnamefont {N.}~\bibnamefont {Wu}},\ }\href
  {https://doi.org/10.1016/j.nuclphysa.2003.12.021} {\bibfield  {journal}
  {\bibinfo  {journal} {Nucl. Phys. A}\ }\textbf {\bibinfo {volume} {733}},\
  \bibinfo {pages} {235} (\bibinfo {year} {2004}{\natexlab{a}})},\ \Eprint
  {https://arxiv.org/abs/hep-ph/0310293} {arXiv:hep-ph/0310293} \BibitemShut
  {NoStop}%
\bibitem [{\citenamefont {Zhou}\ \emph {et~al.}(2005)\citenamefont {Zhou},
  \citenamefont {Qin}, \citenamefont {Zhang}, \citenamefont {Xiao},
  \citenamefont {Zheng},\ and\ \citenamefont {Wu}}]{Zhou:2004ms}%
  \BibitemOpen
  \bibfield  {author} {\bibinfo {author} {\bibfnamefont {Z.~Y.}\ \bibnamefont
  {Zhou}}, \bibinfo {author} {\bibfnamefont {G.~Y.}\ \bibnamefont {Qin}},
  \bibinfo {author} {\bibfnamefont {P.}~\bibnamefont {Zhang}}, \bibinfo
  {author} {\bibfnamefont {Z.}~\bibnamefont {Xiao}}, \bibinfo {author}
  {\bibfnamefont {H.~Q.}\ \bibnamefont {Zheng}},\ and\ \bibinfo {author}
  {\bibfnamefont {N.}~\bibnamefont {Wu}},\ }\href
  {https://doi.org/10.1088/1126-6708/2005/02/043} {\bibfield  {journal}
  {\bibinfo  {journal} {JHEP}\ }\textbf {\bibinfo {volume} {02}},\ \bibinfo
  {pages} {043}},\ \Eprint {https://arxiv.org/abs/hep-ph/0406271}
  {arXiv:hep-ph/0406271} \BibitemShut {NoStop}%
\bibitem [{\citenamefont {Zheng}\ \emph
  {et~al.}(2004{\natexlab{b}})\citenamefont {Zheng}, \citenamefont {Zhou},
  \citenamefont {Qin},\ and\ \citenamefont {Xiao}}]{Zheng:2003rv}%
  \BibitemOpen
  \bibfield  {author} {\bibinfo {author} {\bibfnamefont {H.~Q.}\ \bibnamefont
  {Zheng}}, \bibinfo {author} {\bibfnamefont {Z.~Y.}\ \bibnamefont {Zhou}},
  \bibinfo {author} {\bibfnamefont {G.~Y.}\ \bibnamefont {Qin}},\ and\ \bibinfo
  {author} {\bibfnamefont {Z.}~\bibnamefont {Xiao}},\ }\href
  {https://doi.org/10.1063/1.1799725} {\bibfield  {journal} {\bibinfo
  {journal} {AIP Conf. Proc.}\ }\textbf {\bibinfo {volume} {717}},\ \bibinfo
  {pages} {322} (\bibinfo {year} {2004}{\natexlab{b}})},\ \Eprint
  {https://arxiv.org/abs/hep-ph/0309242} {arXiv:hep-ph/0309242} \BibitemShut
  {NoStop}%
\bibitem [{\citenamefont {Wang}\ \emph {et~al.}(2005)\citenamefont {Wang},
  \citenamefont {Zhou},\ and\ \citenamefont {Zheng}}]{Wang:2005ks}%
  \BibitemOpen
  \bibfield  {author} {\bibinfo {author} {\bibfnamefont {J.-j.}\ \bibnamefont
  {Wang}}, \bibinfo {author} {\bibfnamefont {Z.~Y.}\ \bibnamefont {Zhou}},\
  and\ \bibinfo {author} {\bibfnamefont {H.~Q.}\ \bibnamefont {Zheng}},\ }\href
  {https://doi.org/10.1088/1126-6708/2005/12/019} {\bibfield  {journal}
  {\bibinfo  {journal} {JHEP}\ }\textbf {\bibinfo {volume} {12}},\ \bibinfo
  {pages} {019}},\ \Eprint {https://arxiv.org/abs/hep-ph/0508040}
  {arXiv:hep-ph/0508040} \BibitemShut {NoStop}%
\bibitem [{\citenamefont {Zhou}\ and\ \citenamefont
  {Zheng}(2006)}]{Zhou:2006wm}%
  \BibitemOpen
  \bibfield  {author} {\bibinfo {author} {\bibfnamefont {Z.~Y.}\ \bibnamefont
  {Zhou}}\ and\ \bibinfo {author} {\bibfnamefont {H.~Q.}\ \bibnamefont
  {Zheng}},\ }\href {https://doi.org/10.1016/j.nuclphysa.2006.06.170}
  {\bibfield  {journal} {\bibinfo  {journal} {Nucl. Phys. A}\ }\textbf
  {\bibinfo {volume} {775}},\ \bibinfo {pages} {212} (\bibinfo {year}
  {2006})},\ \Eprint {https://arxiv.org/abs/hep-ph/0603062}
  {arXiv:hep-ph/0603062} \BibitemShut {NoStop}%
\bibitem [{\citenamefont {Roy}(1971)}]{Roy:1971tc}%
  \BibitemOpen
  \bibfield  {author} {\bibinfo {author} {\bibfnamefont {S.~M.}\ \bibnamefont
  {Roy}},\ }\href {https://doi.org/10.1016/0370-2693(71)90724-6} {\bibfield
  {journal} {\bibinfo  {journal} {Phys. Lett. B}\ }\textbf {\bibinfo {volume}
  {36}},\ \bibinfo {pages} {353} (\bibinfo {year} {1971})}\BibitemShut
  {NoStop}%
\bibitem [{\citenamefont {Colangelo}\ \emph {et~al.}(2001)\citenamefont
  {Colangelo}, \citenamefont {Gasser},\ and\ \citenamefont
  {Leutwyler}}]{Colangelo:2001df}%
  \BibitemOpen
  \bibfield  {author} {\bibinfo {author} {\bibfnamefont {G.}~\bibnamefont
  {Colangelo}}, \bibinfo {author} {\bibfnamefont {J.}~\bibnamefont {Gasser}},\
  and\ \bibinfo {author} {\bibfnamefont {H.}~\bibnamefont {Leutwyler}},\ }\href
  {https://doi.org/10.1016/S0550-3213(01)00147-X} {\bibfield  {journal}
  {\bibinfo  {journal} {Nucl. Phys. B}\ }\textbf {\bibinfo {volume} {603}},\
  \bibinfo {pages} {125} (\bibinfo {year} {2001})},\ \Eprint
  {https://arxiv.org/abs/hep-ph/0103088} {arXiv:hep-ph/0103088} \BibitemShut
  {NoStop}%
\bibitem [{\citenamefont {Ananthanarayan}\ \emph {et~al.}(2001)\citenamefont
  {Ananthanarayan}, \citenamefont {Colangelo}, \citenamefont {Gasser},\ and\
  \citenamefont {Leutwyler}}]{Ananthanarayan:2000ht}%
  \BibitemOpen
  \bibfield  {author} {\bibinfo {author} {\bibfnamefont {B.}~\bibnamefont
  {Ananthanarayan}}, \bibinfo {author} {\bibfnamefont {G.}~\bibnamefont
  {Colangelo}}, \bibinfo {author} {\bibfnamefont {J.}~\bibnamefont {Gasser}},\
  and\ \bibinfo {author} {\bibfnamefont {H.}~\bibnamefont {Leutwyler}},\ }\href
  {https://doi.org/10.1016/S0370-1573(01)00009-6} {\bibfield  {journal}
  {\bibinfo  {journal} {Phys. Rept.}\ }\textbf {\bibinfo {volume} {353}},\
  \bibinfo {pages} {207} (\bibinfo {year} {2001})},\ \Eprint
  {https://arxiv.org/abs/hep-ph/0005297} {arXiv:hep-ph/0005297} \BibitemShut
  {NoStop}%
\bibitem [{\citenamefont {Caprini}\ \emph {et~al.}(2006)\citenamefont
  {Caprini}, \citenamefont {Colangelo},\ and\ \citenamefont
  {Leutwyler}}]{Caprini:2004zr}%
  \BibitemOpen
  \bibfield  {author} {\bibinfo {author} {\bibfnamefont {I.}~\bibnamefont
  {Caprini}}, \bibinfo {author} {\bibfnamefont {G.}~\bibnamefont {Colangelo}},\
  and\ \bibinfo {author} {\bibfnamefont {H.}~\bibnamefont {Leutwyler}},\ }\href
  {https://doi.org/10.1103/PhysRevLett.96.132001} {\bibfield  {journal}
  {\bibinfo  {journal} {Phys. Rev. Lett.}\ }\textbf {\bibinfo {volume} {96}},\
  \bibinfo {pages} {132001} (\bibinfo {year} {2006})},\ \Eprint
  {https://arxiv.org/abs/hep-ph/0512364} {arXiv:hep-ph/0512364} \BibitemShut
  {NoStop}%
\bibitem [{\citenamefont {Garcia-Martin}\ \emph {et~al.}(2011)\citenamefont
  {Garcia-Martin}, \citenamefont {Kaminski}, \citenamefont {Pelaez},
  \citenamefont {Ruiz~de Elvira},\ and\ \citenamefont
  {Yndurain}}]{Garcia-Martin:2011iqs}%
  \BibitemOpen
  \bibfield  {author} {\bibinfo {author} {\bibfnamefont {R.}~\bibnamefont
  {Garcia-Martin}}, \bibinfo {author} {\bibfnamefont {R.}~\bibnamefont
  {Kaminski}}, \bibinfo {author} {\bibfnamefont {J.~R.}\ \bibnamefont
  {Pelaez}}, \bibinfo {author} {\bibfnamefont {J.}~\bibnamefont {Ruiz~de
  Elvira}},\ and\ \bibinfo {author} {\bibfnamefont {F.~J.}\ \bibnamefont
  {Yndurain}},\ }\href {https://doi.org/10.1103/PhysRevD.83.074004} {\bibfield
  {journal} {\bibinfo  {journal} {Phys. Rev. D}\ }\textbf {\bibinfo {volume}
  {83}},\ \bibinfo {pages} {074004} (\bibinfo {year} {2011})},\ \Eprint
  {https://arxiv.org/abs/1102.2183} {arXiv:1102.2183 [hep-ph]} \BibitemShut
  {NoStop}%
\bibitem [{\citenamefont {Mennessier}\ \emph {et~al.}(2008)\citenamefont
  {Mennessier}, \citenamefont {Narison},\ and\ \citenamefont
  {Ochs}}]{Mennessier:2008kk}%
  \BibitemOpen
  \bibfield  {author} {\bibinfo {author} {\bibfnamefont {G.}~\bibnamefont
  {Mennessier}}, \bibinfo {author} {\bibfnamefont {S.}~\bibnamefont
  {Narison}},\ and\ \bibinfo {author} {\bibfnamefont {W.}~\bibnamefont
  {Ochs}},\ }\href {https://doi.org/10.1016/j.physletb.2008.06.018} {\bibfield
  {journal} {\bibinfo  {journal} {Phys. Lett. B}\ }\textbf {\bibinfo {volume}
  {665}},\ \bibinfo {pages} {205} (\bibinfo {year} {2008})},\ \Eprint
  {https://arxiv.org/abs/0804.4452} {arXiv:0804.4452 [hep-ph]} \BibitemShut
  {NoStop}%
\bibitem [{\citenamefont {Mennessier}\ \emph {et~al.}(2010)\citenamefont
  {Mennessier}, \citenamefont {Narison},\ and\ \citenamefont
  {Wang}}]{Mennessier:2010xg}%
  \BibitemOpen
  \bibfield  {author} {\bibinfo {author} {\bibfnamefont {G.}~\bibnamefont
  {Mennessier}}, \bibinfo {author} {\bibfnamefont {S.}~\bibnamefont
  {Narison}},\ and\ \bibinfo {author} {\bibfnamefont {X.~G.}\ \bibnamefont
  {Wang}},\ }\href {https://doi.org/10.1016/j.physletb.2010.03.031} {\bibfield
  {journal} {\bibinfo  {journal} {Phys. Lett. B}\ }\textbf {\bibinfo {volume}
  {688}},\ \bibinfo {pages} {59} (\bibinfo {year} {2010})},\ \Eprint
  {https://arxiv.org/abs/1002.1402} {arXiv:1002.1402 [hep-ph]} \BibitemShut
  {NoStop}%
\bibitem [{\citenamefont {Sun}\ \emph {et~al.}(2007)\citenamefont {Sun},
  \citenamefont {Xiao}, \citenamefont {Xiao},\ and\ \citenamefont
  {Zheng}}]{Sun:2005uk}%
  \BibitemOpen
  \bibfield  {author} {\bibinfo {author} {\bibfnamefont {Z.~X.}\ \bibnamefont
  {Sun}}, \bibinfo {author} {\bibfnamefont {L.~Y.}\ \bibnamefont {Xiao}},
  \bibinfo {author} {\bibfnamefont {Z.}~\bibnamefont {Xiao}},\ and\ \bibinfo
  {author} {\bibfnamefont {H.~Q.}\ \bibnamefont {Zheng}},\ }\href
  {https://doi.org/10.1142/S0217732307023304} {\bibfield  {journal} {\bibinfo
  {journal} {Mod. Phys. Lett. A}\ }\textbf {\bibinfo {volume} {22}},\ \bibinfo
  {pages} {711} (\bibinfo {year} {2007})},\ \Eprint
  {https://arxiv.org/abs/hep-ph/0503195} {arXiv:hep-ph/0503195} \BibitemShut
  {NoStop}%
\bibitem [{\citenamefont {Weinberg}(2013)}]{Weinberg:2013cfa}%
  \BibitemOpen
  \bibfield  {author} {\bibinfo {author} {\bibfnamefont {S.}~\bibnamefont
  {Weinberg}},\ }\href {https://doi.org/10.1103/PhysRevLett.110.261601}
  {\bibfield  {journal} {\bibinfo  {journal} {Phys. Rev. Lett.}\ }\textbf
  {\bibinfo {volume} {110}},\ \bibinfo {pages} {261601} (\bibinfo {year}
  {2013})},\ \Eprint {https://arxiv.org/abs/1303.0342} {arXiv:1303.0342
  [hep-ph]} \BibitemShut {NoStop}%
\bibitem [{\citenamefont {Gasser}\ and\ \citenamefont
  {Leutwyler}(1987)}]{Gasser:1987ah}%
  \BibitemOpen
  \bibfield  {author} {\bibinfo {author} {\bibfnamefont {J.}~\bibnamefont
  {Gasser}}\ and\ \bibinfo {author} {\bibfnamefont {H.}~\bibnamefont
  {Leutwyler}},\ }\href {https://doi.org/10.1016/0370-2693(87)91652-2}
  {\bibfield  {journal} {\bibinfo  {journal} {Phys. Lett. B}\ }\textbf
  {\bibinfo {volume} {188}},\ \bibinfo {pages} {477} (\bibinfo {year}
  {1987})}\BibitemShut {NoStop}%
\bibitem [{\citenamefont {Gerber}\ and\ \citenamefont
  {Leutwyler}(1989)}]{Gerber:1988tt}%
  \BibitemOpen
  \bibfield  {author} {\bibinfo {author} {\bibfnamefont {P.}~\bibnamefont
  {Gerber}}\ and\ \bibinfo {author} {\bibfnamefont {H.}~\bibnamefont
  {Leutwyler}},\ }\href {https://doi.org/10.1016/0550-3213(89)90349-0}
  {\bibfield  {journal} {\bibinfo  {journal} {Nucl. Phys. B}\ }\textbf
  {\bibinfo {volume} {321}},\ \bibinfo {pages} {387} (\bibinfo {year}
  {1989})}\BibitemShut {NoStop}%
\bibitem [{\citenamefont {Cort\'es}\ \emph {et~al.}(2016)\citenamefont
  {Cort\'es}, \citenamefont {G\'omez~Nicola},\ and\ \citenamefont
  {Morales}}]{Cortes:2015emo}%
  \BibitemOpen
  \bibfield  {author} {\bibinfo {author} {\bibfnamefont {S.}~\bibnamefont
  {Cort\'es}}, \bibinfo {author} {\bibfnamefont {A.}~\bibnamefont
  {G\'omez~Nicola}},\ and\ \bibinfo {author} {\bibfnamefont {J.}~\bibnamefont
  {Morales}},\ }\href {https://doi.org/10.1103/PhysRevD.93.036001} {\bibfield
  {journal} {\bibinfo  {journal} {Phys. Rev. D}\ }\textbf {\bibinfo {volume}
  {93}},\ \bibinfo {pages} {036001} (\bibinfo {year} {2016})},\ \Eprint
  {https://arxiv.org/abs/1511.00031} {arXiv:1511.00031 [hep-ph]} \BibitemShut
  {NoStop}%
\bibitem [{\citenamefont {Gomez~Nicola}\ and\ \citenamefont {Ruiz~de
  Elvira}(2018)}]{GomezNicola:2017bhm}%
  \BibitemOpen
  \bibfield  {author} {\bibinfo {author} {\bibfnamefont {A.}~\bibnamefont
  {Gomez~Nicola}}\ and\ \bibinfo {author} {\bibfnamefont {J.}~\bibnamefont
  {Ruiz~de Elvira}},\ }\href {https://doi.org/10.1103/PhysRevD.97.074016}
  {\bibfield  {journal} {\bibinfo  {journal} {Phys. Rev. D}\ }\textbf {\bibinfo
  {volume} {97}},\ \bibinfo {pages} {074016} (\bibinfo {year} {2018})},\
  \Eprint {https://arxiv.org/abs/1704.05036} {arXiv:1704.05036 [hep-ph]}
  \BibitemShut {NoStop}%
\bibitem [{\citenamefont {Gao}\ \emph {et~al.}(2019)\citenamefont {Gao},
  \citenamefont {Guo},\ and\ \citenamefont {Pang}}]{Gao:2019idb}%
  \BibitemOpen
  \bibfield  {author} {\bibinfo {author} {\bibfnamefont {R.}~\bibnamefont
  {Gao}}, \bibinfo {author} {\bibfnamefont {Z.-H.}\ \bibnamefont {Guo}},\ and\
  \bibinfo {author} {\bibfnamefont {J.-Y.}\ \bibnamefont {Pang}},\ }\href
  {https://doi.org/10.1103/PhysRevD.100.114028} {\bibfield  {journal} {\bibinfo
   {journal} {Phys. Rev. D}\ }\textbf {\bibinfo {volume} {100}},\ \bibinfo
  {pages} {114028} (\bibinfo {year} {2019})},\ \Eprint
  {https://arxiv.org/abs/1907.01787} {arXiv:1907.01787 [hep-ph]} \BibitemShut
  {NoStop}%
\bibitem [{\citenamefont {Dolan}\ and\ \citenamefont
  {Jackiw}(1974)}]{Dolan:1973qd}%
  \BibitemOpen
  \bibfield  {author} {\bibinfo {author} {\bibfnamefont {L.}~\bibnamefont
  {Dolan}}\ and\ \bibinfo {author} {\bibfnamefont {R.}~\bibnamefont {Jackiw}},\
  }\href {https://doi.org/10.1103/PhysRevD.9.3320} {\bibfield  {journal}
  {\bibinfo  {journal} {Phys. Rev. D}\ }\textbf {\bibinfo {volume} {9}},\
  \bibinfo {pages} {3320} (\bibinfo {year} {1974})}\BibitemShut {NoStop}%
\bibitem [{\citenamefont {Schnitzer}(1974)}]{Schnitzer:1974ji}%
  \BibitemOpen
  \bibfield  {author} {\bibinfo {author} {\bibfnamefont {H.~J.}\ \bibnamefont
  {Schnitzer}},\ }\href {https://doi.org/10.1103/PhysRevD.10.1800} {\bibfield
  {journal} {\bibinfo  {journal} {Phys. Rev. D}\ }\textbf {\bibinfo {volume}
  {10}},\ \bibinfo {pages} {1800} (\bibinfo {year} {1974})}\BibitemShut
  {NoStop}%
\bibitem [{\citenamefont {Coleman}\ \emph {et~al.}(1974)\citenamefont
  {Coleman}, \citenamefont {Jackiw},\ and\ \citenamefont
  {Politzer}}]{Coleman:1974jh}%
  \BibitemOpen
  \bibfield  {author} {\bibinfo {author} {\bibfnamefont {S.~R.}\ \bibnamefont
  {Coleman}}, \bibinfo {author} {\bibfnamefont {R.}~\bibnamefont {Jackiw}},\
  and\ \bibinfo {author} {\bibfnamefont {H.~D.}\ \bibnamefont {Politzer}},\
  }\href {https://doi.org/10.1103/PhysRevD.10.2491} {\bibfield  {journal}
  {\bibinfo  {journal} {Phys. Rev. D}\ }\textbf {\bibinfo {volume} {10}},\
  \bibinfo {pages} {2491} (\bibinfo {year} {1974})}\BibitemShut {NoStop}%
\bibitem [{\citenamefont {Guo}\ \emph {et~al.}(2007{\natexlab{b}})\citenamefont
  {Guo}, \citenamefont {Xiao},\ and\ \citenamefont {Zheng}}]{Guo:2006br}%
  \BibitemOpen
  \bibfield  {author} {\bibinfo {author} {\bibfnamefont {Z.-H.}\ \bibnamefont
  {Guo}}, \bibinfo {author} {\bibfnamefont {L.~Y.}\ \bibnamefont {Xiao}},\ and\
  \bibinfo {author} {\bibfnamefont {H.~Q.}\ \bibnamefont {Zheng}},\ }\href
  {https://doi.org/10.1142/S0217751X0703710X} {\bibfield  {journal} {\bibinfo
  {journal} {Int. J. Mod. Phys. A}\ }\textbf {\bibinfo {volume} {22}},\
  \bibinfo {pages} {4603} (\bibinfo {year} {2007}{\natexlab{b}})},\ \Eprint
  {https://arxiv.org/abs/hep-ph/0610434} {arXiv:hep-ph/0610434} \BibitemShut
  {NoStop}%
\bibitem [{\citenamefont {Luscher}(1986)}]{Luscher:1986pf}%
  \BibitemOpen
  \bibfield  {author} {\bibinfo {author} {\bibfnamefont {M.}~\bibnamefont
  {Luscher}},\ }\href {https://doi.org/10.1007/BF01211097} {\bibfield
  {journal} {\bibinfo  {journal} {Commun. Math. Phys.}\ }\textbf {\bibinfo
  {volume} {105}},\ \bibinfo {pages} {153} (\bibinfo {year}
  {1986})}\BibitemShut {NoStop}%
\bibitem [{\citenamefont {Luscher}\ and\ \citenamefont
  {Wolff}(1990)}]{Luscher:1990ck}%
  \BibitemOpen
  \bibfield  {author} {\bibinfo {author} {\bibfnamefont {M.}~\bibnamefont
  {Luscher}}\ and\ \bibinfo {author} {\bibfnamefont {U.}~\bibnamefont
  {Wolff}},\ }\href {https://doi.org/10.1016/0550-3213(90)90540-T} {\bibfield
  {journal} {\bibinfo  {journal} {Nucl. Phys. B}\ }\textbf {\bibinfo {volume}
  {339}},\ \bibinfo {pages} {222} (\bibinfo {year} {1990})}\BibitemShut
  {NoStop}%
\bibitem [{\citenamefont {Luscher}(1991)}]{Luscher:1990ux}%
  \BibitemOpen
  \bibfield  {author} {\bibinfo {author} {\bibfnamefont {M.}~\bibnamefont
  {Luscher}},\ }\href {https://doi.org/10.1016/0550-3213(91)90366-6} {\bibfield
   {journal} {\bibinfo  {journal} {Nucl. Phys. B}\ }\textbf {\bibinfo {volume}
  {354}},\ \bibinfo {pages} {531} (\bibinfo {year} {1991})}\BibitemShut
  {NoStop}%
\bibitem [{\citenamefont {Kuramashi}\ \emph {et~al.}(1993)\citenamefont
  {Kuramashi}, \citenamefont {Fukugita}, \citenamefont {Mino}, \citenamefont
  {Okawa},\ and\ \citenamefont {Ukawa}}]{Kuramashi:1993ka}%
  \BibitemOpen
  \bibfield  {author} {\bibinfo {author} {\bibfnamefont {Y.}~\bibnamefont
  {Kuramashi}}, \bibinfo {author} {\bibfnamefont {M.}~\bibnamefont {Fukugita}},
  \bibinfo {author} {\bibfnamefont {H.}~\bibnamefont {Mino}}, \bibinfo {author}
  {\bibfnamefont {M.}~\bibnamefont {Okawa}},\ and\ \bibinfo {author}
  {\bibfnamefont {A.}~\bibnamefont {Ukawa}},\ }\href
  {https://doi.org/10.1103/PhysRevLett.71.2387} {\bibfield  {journal} {\bibinfo
   {journal} {Phys. Rev. Lett.}\ }\textbf {\bibinfo {volume} {71}},\ \bibinfo
  {pages} {2387} (\bibinfo {year} {1993})}\BibitemShut {NoStop}%
\bibitem [{\citenamefont {He}\ \emph {et~al.}(2005)\citenamefont {He},
  \citenamefont {Feng},\ and\ \citenamefont {Liu}}]{He:2005ey}%
  \BibitemOpen
  \bibfield  {author} {\bibinfo {author} {\bibfnamefont {S.}~\bibnamefont
  {He}}, \bibinfo {author} {\bibfnamefont {X.}~\bibnamefont {Feng}},\ and\
  \bibinfo {author} {\bibfnamefont {C.}~\bibnamefont {Liu}},\ }\href
  {https://doi.org/10.1088/1126-6708/2005/07/011} {\bibfield  {journal}
  {\bibinfo  {journal} {JHEP}\ }\textbf {\bibinfo {volume} {07}},\ \bibinfo
  {pages} {011}},\ \Eprint {https://arxiv.org/abs/hep-lat/0504019}
  {arXiv:hep-lat/0504019} \BibitemShut {NoStop}%
\bibitem [{\citenamefont {Mathur}\ \emph {et~al.}(2007)\citenamefont {Mathur},
  \citenamefont {Alexandru}, \citenamefont {Chen}, \citenamefont {Dong},
  \citenamefont {Draper}, \citenamefont {Horvath}, \citenamefont {Lee},
  \citenamefont {Liu}, \citenamefont {Tamhankar},\ and\ \citenamefont
  {Zhang}}]{Mathur:2006bs}%
  \BibitemOpen
  \bibfield  {author} {\bibinfo {author} {\bibfnamefont {N.}~\bibnamefont
  {Mathur}}, \bibinfo {author} {\bibfnamefont {A.}~\bibnamefont {Alexandru}},
  \bibinfo {author} {\bibfnamefont {Y.}~\bibnamefont {Chen}}, \bibinfo {author}
  {\bibfnamefont {S.~J.}\ \bibnamefont {Dong}}, \bibinfo {author}
  {\bibfnamefont {T.}~\bibnamefont {Draper}}, \bibinfo {author} {\bibfnamefont
  {I.}~\bibnamefont {Horvath}}, \bibinfo {author} {\bibfnamefont {F.~X.}\
  \bibnamefont {Lee}}, \bibinfo {author} {\bibfnamefont {K.~F.}\ \bibnamefont
  {Liu}}, \bibinfo {author} {\bibfnamefont {S.}~\bibnamefont {Tamhankar}},\
  and\ \bibinfo {author} {\bibfnamefont {J.~B.}\ \bibnamefont {Zhang}},\ }\href
  {https://doi.org/10.1103/PhysRevD.76.114505} {\bibfield  {journal} {\bibinfo
  {journal} {Phys. Rev. D}\ }\textbf {\bibinfo {volume} {76}},\ \bibinfo
  {pages} {114505} (\bibinfo {year} {2007})},\ \Eprint
  {https://arxiv.org/abs/hep-ph/0607110} {arXiv:hep-ph/0607110} \BibitemShut
  {NoStop}%
\bibitem [{\citenamefont {Feng}\ \emph {et~al.}(2010)\citenamefont {Feng},
  \citenamefont {Jansen},\ and\ \citenamefont {Renner}}]{Feng:2009ij}%
  \BibitemOpen
  \bibfield  {author} {\bibinfo {author} {\bibfnamefont {X.}~\bibnamefont
  {Feng}}, \bibinfo {author} {\bibfnamefont {K.}~\bibnamefont {Jansen}},\ and\
  \bibinfo {author} {\bibfnamefont {D.~B.}\ \bibnamefont {Renner}},\ }\href
  {https://doi.org/10.1016/j.physletb.2010.01.018} {\bibfield  {journal}
  {\bibinfo  {journal} {Phys. Lett. B}\ }\textbf {\bibinfo {volume} {684}},\
  \bibinfo {pages} {268} (\bibinfo {year} {2010})},\ \Eprint
  {https://arxiv.org/abs/0909.3255} {arXiv:0909.3255 [hep-lat]} \BibitemShut
  {NoStop}%
\bibitem [{\citenamefont {Fu}(2012)}]{Fu:2011xz}%
  \BibitemOpen
  \bibfield  {author} {\bibinfo {author} {\bibfnamefont {Z.}~\bibnamefont
  {Fu}},\ }\href {https://doi.org/10.1103/PhysRevD.85.014506} {\bibfield
  {journal} {\bibinfo  {journal} {Phys. Rev. D}\ }\textbf {\bibinfo {volume}
  {85}},\ \bibinfo {pages} {014506} (\bibinfo {year} {2012})},\ \Eprint
  {https://arxiv.org/abs/1110.0319} {arXiv:1110.0319 [hep-lat]} \BibitemShut
  {NoStop}%
\bibitem [{\citenamefont {Briceno}\ \emph {et~al.}(2017)\citenamefont
  {Briceno}, \citenamefont {Dudek}, \citenamefont {Edwards},\ and\
  \citenamefont {Wilson}}]{Briceno:2016mjc}%
  \BibitemOpen
  \bibfield  {author} {\bibinfo {author} {\bibfnamefont {R.~A.}\ \bibnamefont
  {Briceno}}, \bibinfo {author} {\bibfnamefont {J.~J.}\ \bibnamefont {Dudek}},
  \bibinfo {author} {\bibfnamefont {R.~G.}\ \bibnamefont {Edwards}},\ and\
  \bibinfo {author} {\bibfnamefont {D.~J.}\ \bibnamefont {Wilson}},\ }\href
  {https://doi.org/10.1103/PhysRevLett.118.022002} {\bibfield  {journal}
  {\bibinfo  {journal} {Phys. Rev. Lett.}\ }\textbf {\bibinfo {volume} {118}},\
  \bibinfo {pages} {022002} (\bibinfo {year} {2017})},\ \Eprint
  {https://arxiv.org/abs/1607.05900} {arXiv:1607.05900 [hep-ph]} \BibitemShut
  {NoStop}%
\bibitem [{\citenamefont {Lin}\ \emph {et~al.}(2009)\citenamefont {Lin} \emph
  {et~al.}}]{HadronSpectrum:2008xlg}%
  \BibitemOpen
  \bibfield  {author} {\bibinfo {author} {\bibfnamefont {H.-W.}\ \bibnamefont
  {Lin}} \emph {et~al.} (\bibinfo {collaboration} {Hadron Spectrum}),\ }\href
  {https://doi.org/10.1103/PhysRevD.79.034502} {\bibfield  {journal} {\bibinfo
  {journal} {Phys. Rev. D}\ }\textbf {\bibinfo {volume} {79}},\ \bibinfo
  {pages} {034502} (\bibinfo {year} {2009})},\ \Eprint
  {https://arxiv.org/abs/0810.3588} {arXiv:0810.3588 [hep-lat]} \BibitemShut
  {NoStop}%
\bibitem [{\citenamefont {Rodas}\ \emph {et~al.}(2023)\citenamefont {Rodas},
  \citenamefont {Dudek},\ and\ \citenamefont {Edwards}}]{Rodas:2023gma}%
  \BibitemOpen
  \bibfield  {author} {\bibinfo {author} {\bibfnamefont {A.}~\bibnamefont
  {Rodas}}, \bibinfo {author} {\bibfnamefont {J.~J.}\ \bibnamefont {Dudek}},\
  and\ \bibinfo {author} {\bibfnamefont {R.~G.}\ \bibnamefont {Edwards}}
  (\bibinfo {collaboration} {Hadron Spectrum}),\ }\href
  {https://doi.org/10.1103/PhysRevD.108.034513} {\bibfield  {journal} {\bibinfo
   {journal} {Phys. Rev. D}\ }\textbf {\bibinfo {volume} {108}},\ \bibinfo
  {pages} {034513} (\bibinfo {year} {2023})},\ \Eprint
  {https://arxiv.org/abs/2303.10701} {arXiv:2303.10701 [hep-lat]} \BibitemShut
  {NoStop}%
\bibitem [{\citenamefont {Hanhart}\ \emph {et~al.}(2008)\citenamefont
  {Hanhart}, \citenamefont {Pelaez},\ and\ \citenamefont
  {Rios}}]{Hanhart:2008mx}%
  \BibitemOpen
  \bibfield  {author} {\bibinfo {author} {\bibfnamefont {C.}~\bibnamefont
  {Hanhart}}, \bibinfo {author} {\bibfnamefont {J.~R.}\ \bibnamefont
  {Pelaez}},\ and\ \bibinfo {author} {\bibfnamefont {G.}~\bibnamefont {Rios}},\
  }\href {https://doi.org/10.1103/PhysRevLett.100.152001} {\bibfield  {journal}
  {\bibinfo  {journal} {Phys. Rev. Lett.}\ }\textbf {\bibinfo {volume} {100}},\
  \bibinfo {pages} {152001} (\bibinfo {year} {2008})},\ \Eprint
  {https://arxiv.org/abs/0801.2871} {arXiv:0801.2871 [hep-ph]} \BibitemShut
  {NoStop}%
\bibitem [{\citenamefont {Pelaez}\ and\ \citenamefont
  {Rios}(2010)}]{Pelaez:2010fj}%
  \BibitemOpen
  \bibfield  {author} {\bibinfo {author} {\bibfnamefont {J.~R.}\ \bibnamefont
  {Pelaez}}\ and\ \bibinfo {author} {\bibfnamefont {G.}~\bibnamefont {Rios}},\
  }\href {https://doi.org/10.1103/PhysRevD.82.114002} {\bibfield  {journal}
  {\bibinfo  {journal} {Phys. Rev. D}\ }\textbf {\bibinfo {volume} {82}},\
  \bibinfo {pages} {114002} (\bibinfo {year} {2010})},\ \Eprint
  {https://arxiv.org/abs/1010.6008} {arXiv:1010.6008 [hep-ph]} \BibitemShut
  {NoStop}%
\bibitem [{\citenamefont {Hanhart}\ \emph {et~al.}(2014)\citenamefont
  {Hanhart}, \citenamefont {Pelaez},\ and\ \citenamefont
  {Rios}}]{Hanhart:2014ssa}%
  \BibitemOpen
  \bibfield  {author} {\bibinfo {author} {\bibfnamefont {C.}~\bibnamefont
  {Hanhart}}, \bibinfo {author} {\bibfnamefont {J.~R.}\ \bibnamefont
  {Pelaez}},\ and\ \bibinfo {author} {\bibfnamefont {G.}~\bibnamefont {Rios}},\
  }\href {https://doi.org/10.1016/j.physletb.2014.11.011} {\bibfield  {journal}
  {\bibinfo  {journal} {Phys. Lett. B}\ }\textbf {\bibinfo {volume} {739}},\
  \bibinfo {pages} {375} (\bibinfo {year} {2014})},\ \Eprint
  {https://arxiv.org/abs/1407.7452} {arXiv:1407.7452 [hep-ph]} \BibitemShut
  {NoStop}%
\bibitem [{\citenamefont {Gao}\ \emph {et~al.}(2022)\citenamefont {Gao},
  \citenamefont {Guo}, \citenamefont {Xiao},\ and\ \citenamefont
  {Zhou}}]{Gao:2022dln}%
  \BibitemOpen
  \bibfield  {author} {\bibinfo {author} {\bibfnamefont {X.-L.}\ \bibnamefont
  {Gao}}, \bibinfo {author} {\bibfnamefont {Z.-H.}\ \bibnamefont {Guo}},
  \bibinfo {author} {\bibfnamefont {Z.}~\bibnamefont {Xiao}},\ and\ \bibinfo
  {author} {\bibfnamefont {Z.-Y.}\ \bibnamefont {Zhou}},\ }\href
  {https://doi.org/10.1103/PhysRevD.105.094002} {\bibfield  {journal} {\bibinfo
   {journal} {Phys. Rev. D}\ }\textbf {\bibinfo {volume} {105}},\ \bibinfo
  {pages} {094002} (\bibinfo {year} {2022})},\ \Eprint
  {https://arxiv.org/abs/2202.03124} {arXiv:2202.03124 [hep-ph]} \BibitemShut
  {NoStop}%
\bibitem [{\citenamefont {Gao}\ \emph {et~al.}(2023)\citenamefont {Gao},
  \citenamefont {Guo}, \citenamefont {Xiao},\ and\ \citenamefont
  {Zhou}}]{Gao:2022tlh}%
  \BibitemOpen
  \bibfield  {author} {\bibinfo {author} {\bibfnamefont {X.-L.}\ \bibnamefont
  {Gao}}, \bibinfo {author} {\bibfnamefont {Z.-H.}\ \bibnamefont {Guo}},
  \bibinfo {author} {\bibfnamefont {Z.}~\bibnamefont {Xiao}},\ and\ \bibinfo
  {author} {\bibfnamefont {Z.-Y.}\ \bibnamefont {Zhou}},\ }\href
  {https://doi.org/10.1103/PhysRevD.107.058502} {\bibfield  {journal} {\bibinfo
   {journal} {Phys. Rev. D}\ }\textbf {\bibinfo {volume} {107}},\ \bibinfo
  {pages} {058502} (\bibinfo {year} {2023})},\ \Eprint
  {https://arxiv.org/abs/2204.01562} {arXiv:2204.01562 [hep-ph]} \BibitemShut
  {NoStop}%
\bibitem [{\citenamefont {Cao}\ \emph {et~al.}(2023)\citenamefont {Cao},
  \citenamefont {Li}, \citenamefont {Guo},\ and\ \citenamefont
  {Zheng}}]{Cao:2023ntr}%
  \BibitemOpen
  \bibfield  {author} {\bibinfo {author} {\bibfnamefont {X.-H.}\ \bibnamefont
  {Cao}}, \bibinfo {author} {\bibfnamefont {Q.-Z.}\ \bibnamefont {Li}},
  \bibinfo {author} {\bibfnamefont {Z.-H.}\ \bibnamefont {Guo}},\ and\ \bibinfo
  {author} {\bibfnamefont {H.-Q.}\ \bibnamefont {Zheng}},\ }\href
  {https://doi.org/10.1103/PhysRevD.108.034009} {\bibfield  {journal} {\bibinfo
   {journal} {Phys. Rev. D}\ }\textbf {\bibinfo {volume} {108}},\ \bibinfo
  {pages} {034009} (\bibinfo {year} {2023})},\ \Eprint
  {https://arxiv.org/abs/2303.02596} {arXiv:2303.02596 [hep-ph]} \BibitemShut
  {NoStop}%
\bibitem [{\citenamefont {Rodas}\ \emph {et~al.}(2024)\citenamefont {Rodas},
  \citenamefont {Dudek},\ and\ \citenamefont {Edwards}}]{Rodas:2024qhn}%
  \BibitemOpen
  \bibfield  {author} {\bibinfo {author} {\bibfnamefont {A.}~\bibnamefont
  {Rodas}}, \bibinfo {author} {\bibfnamefont {J.~J.}\ \bibnamefont {Dudek}},\
  and\ \bibinfo {author} {\bibfnamefont {R.~G.}\ \bibnamefont {Edwards}}
  (\bibinfo {collaboration} {Hadron Spectrum}),\ }\href
  {https://doi.org/10.1103/PhysRevD.109.034513} {\bibfield  {journal} {\bibinfo
   {journal} {Phys. Rev. D}\ }\textbf {\bibinfo {volume} {109}},\ \bibinfo
  {pages} {034513} (\bibinfo {year} {2024})}\BibitemShut {NoStop}%
\bibitem [{\citenamefont {Chivukula}\ and\ \citenamefont
  {Golden}(1991)}]{Chivukula:1991bx}%
  \BibitemOpen
  \bibfield  {author} {\bibinfo {author} {\bibfnamefont {R.~S.}\ \bibnamefont
  {Chivukula}}\ and\ \bibinfo {author} {\bibfnamefont {M.}~\bibnamefont
  {Golden}},\ }\href {https://doi.org/10.1016/0370-2693(91)91253-R} {\bibfield
  {journal} {\bibinfo  {journal} {Phys. Lett. B}\ }\textbf {\bibinfo {volume}
  {267}},\ \bibinfo {pages} {233} (\bibinfo {year} {1991})}\BibitemShut
  {NoStop}%
\bibitem [{\citenamefont {Adler}(1965)}]{Adler:1964um}%
  \BibitemOpen
  \bibfield  {author} {\bibinfo {author} {\bibfnamefont {S.~L.}\ \bibnamefont
  {Adler}},\ }\href {https://doi.org/10.1103/PhysRev.137.B1022} {\bibfield
  {journal} {\bibinfo  {journal} {Phys. Rev.}\ }\textbf {\bibinfo {volume}
  {137}},\ \bibinfo {pages} {B1022} (\bibinfo {year} {1965})}\BibitemShut
  {NoStop}%
\bibitem [{\citenamefont {Gomez~Nicola}\ \emph {et~al.}(2008)\citenamefont
  {Gomez~Nicola}, \citenamefont {Pelaez},\ and\ \citenamefont
  {Rios}}]{GomezNicola:2007qj}%
  \BibitemOpen
  \bibfield  {author} {\bibinfo {author} {\bibfnamefont {A.}~\bibnamefont
  {Gomez~Nicola}}, \bibinfo {author} {\bibfnamefont {J.~R.}\ \bibnamefont
  {Pelaez}},\ and\ \bibinfo {author} {\bibfnamefont {G.}~\bibnamefont {Rios}},\
  }\href {https://doi.org/10.1103/PhysRevD.77.056006} {\bibfield  {journal}
  {\bibinfo  {journal} {Phys. Rev. D}\ }\textbf {\bibinfo {volume} {77}},\
  \bibinfo {pages} {056006} (\bibinfo {year} {2008})},\ \Eprint
  {https://arxiv.org/abs/0712.2763} {arXiv:0712.2763 [hep-ph]} \BibitemShut
  {NoStop}%
\bibitem [{\citenamefont {Chew}\ and\ \citenamefont
  {Mandelstam}(1960)}]{Chew:1960iv}%
  \BibitemOpen
  \bibfield  {author} {\bibinfo {author} {\bibfnamefont {G.~F.}\ \bibnamefont
  {Chew}}\ and\ \bibinfo {author} {\bibfnamefont {S.}~\bibnamefont
  {Mandelstam}},\ }\href {https://doi.org/10.1103/PhysRev.119.467} {\bibfield
  {journal} {\bibinfo  {journal} {Phys. Rev.}\ }\textbf {\bibinfo {volume}
  {119}},\ \bibinfo {pages} {467} (\bibinfo {year} {1960})}\BibitemShut
  {NoStop}%
\bibitem [{\citenamefont {Castillejo}\ \emph {et~al.}(1956)\citenamefont
  {Castillejo}, \citenamefont {Dalitz},\ and\ \citenamefont
  {Dyson}}]{Castillejo:1955ed}%
  \BibitemOpen
  \bibfield  {author} {\bibinfo {author} {\bibfnamefont {L.}~\bibnamefont
  {Castillejo}}, \bibinfo {author} {\bibfnamefont {R.~H.}\ \bibnamefont
  {Dalitz}},\ and\ \bibinfo {author} {\bibfnamefont {F.~J.}\ \bibnamefont
  {Dyson}},\ }\href {https://doi.org/10.1103/PhysRev.101.453} {\bibfield
  {journal} {\bibinfo  {journal} {Phys. Rev.}\ }\textbf {\bibinfo {volume}
  {101}},\ \bibinfo {pages} {453} (\bibinfo {year} {1956})}\BibitemShut
  {NoStop}%
\bibitem [{\citenamefont {Taylor}(1972)}]{Taylor:1972pty}%
  \BibitemOpen
  \bibfield  {author} {\bibinfo {author} {\bibfnamefont {J.~R.}\ \bibnamefont
  {Taylor}},\ }\href@noop {} {\emph {\bibinfo {title} {{Scattering Theory: The
  Quantum Theory of Nonrelativistic Collisions}}}}\ (\bibinfo  {publisher}
  {John Wiley \& Sons, Inc.},\ \bibinfo {address} {New York},\ \bibinfo {year}
  {1972})\BibitemShut {NoStop}%
\bibitem [{\citenamefont {Pisarski}\ and\ \citenamefont
  {Wilczek}(1984)}]{Pisarski:1983ms}%
  \BibitemOpen
  \bibfield  {author} {\bibinfo {author} {\bibfnamefont {R.~D.}\ \bibnamefont
  {Pisarski}}\ and\ \bibinfo {author} {\bibfnamefont {F.}~\bibnamefont
  {Wilczek}},\ }\href {https://doi.org/10.1103/PhysRevD.29.338} {\bibfield
  {journal} {\bibinfo  {journal} {Phys. Rev. D}\ }\textbf {\bibinfo {volume}
  {29}},\ \bibinfo {pages} {338} (\bibinfo {year} {1984})}\BibitemShut
  {NoStop}%
\bibitem [{\citenamefont {Bazavov}\ \emph {et~al.}(2012)\citenamefont {Bazavov}
  \emph {et~al.}}]{Bazavov:2011nk}%
  \BibitemOpen
  \bibfield  {author} {\bibinfo {author} {\bibfnamefont {A.}~\bibnamefont
  {Bazavov}} \emph {et~al.},\ }\href
  {https://doi.org/10.1103/PhysRevD.85.054503} {\bibfield  {journal} {\bibinfo
  {journal} {Phys. Rev. D}\ }\textbf {\bibinfo {volume} {85}},\ \bibinfo
  {pages} {054503} (\bibinfo {year} {2012})},\ \Eprint
  {https://arxiv.org/abs/1111.1710} {arXiv:1111.1710 [hep-lat]} \BibitemShut
  {NoStop}%
\bibitem [{\citenamefont {Bazavov}\ \emph {et~al.}(2019)\citenamefont {Bazavov}
  \emph {et~al.}}]{HotQCD:2018pds}%
  \BibitemOpen
  \bibfield  {author} {\bibinfo {author} {\bibfnamefont {A.}~\bibnamefont
  {Bazavov}} \emph {et~al.} (\bibinfo {collaboration} {HotQCD}),\ }\href
  {https://doi.org/10.1016/j.physletb.2019.05.013} {\bibfield  {journal}
  {\bibinfo  {journal} {Phys. Lett. B}\ }\textbf {\bibinfo {volume} {795}},\
  \bibinfo {pages} {15} (\bibinfo {year} {2019})},\ \Eprint
  {https://arxiv.org/abs/1812.08235} {arXiv:1812.08235 [hep-lat]} \BibitemShut
  {NoStop}%
\bibitem [{\citenamefont {Ding}\ \emph {et~al.}(2019)\citenamefont {Ding} \emph
  {et~al.}}]{HotQCD:2019xnw}%
  \BibitemOpen
  \bibfield  {author} {\bibinfo {author} {\bibfnamefont {H.~T.}\ \bibnamefont
  {Ding}} \emph {et~al.} (\bibinfo {collaboration} {HotQCD}),\ }\href
  {https://doi.org/10.1103/PhysRevLett.123.062002} {\bibfield  {journal}
  {\bibinfo  {journal} {Phys. Rev. Lett.}\ }\textbf {\bibinfo {volume} {123}},\
  \bibinfo {pages} {062002} (\bibinfo {year} {2019})},\ \Eprint
  {https://arxiv.org/abs/1903.04801} {arXiv:1903.04801 [hep-lat]} \BibitemShut
  {NoStop}%
\bibitem [{\citenamefont {Bochkarev}\ and\ \citenamefont
  {Kapusta}(1996)}]{Bochkarev:1995gi}%
  \BibitemOpen
  \bibfield  {author} {\bibinfo {author} {\bibfnamefont {A.}~\bibnamefont
  {Bochkarev}}\ and\ \bibinfo {author} {\bibfnamefont {J.~I.}\ \bibnamefont
  {Kapusta}},\ }\href {https://doi.org/10.1103/PhysRevD.54.4066} {\bibfield
  {journal} {\bibinfo  {journal} {Phys. Rev. D}\ }\textbf {\bibinfo {volume}
  {54}},\ \bibinfo {pages} {4066} (\bibinfo {year} {1996})},\ \Eprint
  {https://arxiv.org/abs/hep-ph/9602405} {arXiv:hep-ph/9602405} \BibitemShut
  {NoStop}%
\bibitem [{\citenamefont {Andersen}\ \emph {et~al.}(2004)\citenamefont
  {Andersen}, \citenamefont {Boer},\ and\ \citenamefont
  {Warringa}}]{Andersen:2004ae}%
  \BibitemOpen
  \bibfield  {author} {\bibinfo {author} {\bibfnamefont {J.~O.}\ \bibnamefont
  {Andersen}}, \bibinfo {author} {\bibfnamefont {D.}~\bibnamefont {Boer}},\
  and\ \bibinfo {author} {\bibfnamefont {H.~J.}\ \bibnamefont {Warringa}},\
  }\href {https://doi.org/10.1103/PhysRevD.70.116007} {\bibfield  {journal}
  {\bibinfo  {journal} {Phys. Rev. D}\ }\textbf {\bibinfo {volume} {70}},\
  \bibinfo {pages} {116007} (\bibinfo {year} {2004})},\ \Eprint
  {https://arxiv.org/abs/hep-ph/0408033} {arXiv:hep-ph/0408033} \BibitemShut
  {NoStop}%
\bibitem [{\citenamefont {Meyers-Ortmanns}\ \emph {et~al.}(1993)\citenamefont
  {Meyers-Ortmanns}, \citenamefont {Pirner},\ and\ \citenamefont
  {Schaefer}}]{Meyers-Ortmanns:1993dhx}%
  \BibitemOpen
  \bibfield  {author} {\bibinfo {author} {\bibfnamefont {H.}~\bibnamefont
  {Meyers-Ortmanns}}, \bibinfo {author} {\bibfnamefont {H.~J.}\ \bibnamefont
  {Pirner}},\ and\ \bibinfo {author} {\bibfnamefont {B.~J.}\ \bibnamefont
  {Schaefer}},\ }\href {https://doi.org/10.1016/0370-2693(93)90557-X}
  {\bibfield  {journal} {\bibinfo  {journal} {Phys. Lett. B}\ }\textbf
  {\bibinfo {volume} {311}},\ \bibinfo {pages} {213} (\bibinfo {year}
  {1993})}\BibitemShut {NoStop}%
\bibitem [{\citenamefont {Meyer-Ortmanns}(1996)}]{Meyer-Ortmanns:1996ioo}%
  \BibitemOpen
  \bibfield  {author} {\bibinfo {author} {\bibfnamefont {H.}~\bibnamefont
  {Meyer-Ortmanns}},\ }\href {https://doi.org/10.1103/RevModPhys.68.473}
  {\bibfield  {journal} {\bibinfo  {journal} {Rev. Mod. Phys.}\ }\textbf
  {\bibinfo {volume} {68}},\ \bibinfo {pages} {473} (\bibinfo {year} {1996})},\
  \Eprint {https://arxiv.org/abs/hep-lat/9608098} {arXiv:hep-lat/9608098}
  \BibitemShut {NoStop}%
\bibitem [{\citenamefont {Bellac}(2011)}]{Bellac:2011kqa}%
  \BibitemOpen
  \bibfield  {author} {\bibinfo {author} {\bibfnamefont {M.~L.}\ \bibnamefont
  {Bellac}},\ }\href {https://doi.org/10.1017/CBO9780511721700} {\emph
  {\bibinfo {title} {{Thermal Field Theory}}}},\ Cambridge Monographs on
  Mathematical Physics\ (\bibinfo  {publisher} {Cambridge University Press},\
  \bibinfo {year} {2011})\BibitemShut {NoStop}%
\bibitem [{\citenamefont {Quack}\ \emph {et~al.}(1995)\citenamefont {Quack},
  \citenamefont {Zhuang}, \citenamefont {Kalinovsky}, \citenamefont
  {Klevansky},\ and\ \citenamefont {Hufner}}]{Quack:1994vc}%
  \BibitemOpen
  \bibfield  {author} {\bibinfo {author} {\bibfnamefont {E.}~\bibnamefont
  {Quack}}, \bibinfo {author} {\bibfnamefont {P.}~\bibnamefont {Zhuang}},
  \bibinfo {author} {\bibfnamefont {Y.}~\bibnamefont {Kalinovsky}}, \bibinfo
  {author} {\bibfnamefont {S.~P.}\ \bibnamefont {Klevansky}},\ and\ \bibinfo
  {author} {\bibfnamefont {J.}~\bibnamefont {Hufner}},\ }\href
  {https://doi.org/10.1016/0370-2693(95)00128-8} {\bibfield  {journal}
  {\bibinfo  {journal} {Phys. Lett. B}\ }\textbf {\bibinfo {volume} {348}},\
  \bibinfo {pages} {1} (\bibinfo {year} {1995})},\ \Eprint
  {https://arxiv.org/abs/hep-ph/9410243} {arXiv:hep-ph/9410243} \BibitemShut
  {NoStop}%
\bibitem [{\citenamefont {Kaiser}(1999)}]{Kaiser:1999mt}%
  \BibitemOpen
  \bibfield  {author} {\bibinfo {author} {\bibfnamefont {N.}~\bibnamefont
  {Kaiser}},\ }\href {https://doi.org/10.1103/PhysRevC.59.2945} {\bibfield
  {journal} {\bibinfo  {journal} {Phys. Rev. C}\ }\textbf {\bibinfo {volume}
  {59}},\ \bibinfo {pages} {2945} (\bibinfo {year} {1999})}\BibitemShut
  {NoStop}%
\bibitem [{\citenamefont {Gomez~Nicola}\ \emph {et~al.}(2002)\citenamefont
  {Gomez~Nicola}, \citenamefont {Llanes-Estrada},\ and\ \citenamefont
  {Pelaez}}]{GomezNicola:2002tn}%
  \BibitemOpen
  \bibfield  {author} {\bibinfo {author} {\bibfnamefont {A.}~\bibnamefont
  {Gomez~Nicola}}, \bibinfo {author} {\bibfnamefont {F.~J.}\ \bibnamefont
  {Llanes-Estrada}},\ and\ \bibinfo {author} {\bibfnamefont {J.~R.}\
  \bibnamefont {Pelaez}},\ }\href
  {https://doi.org/10.1016/S0370-2693(02)02959-3} {\bibfield  {journal}
  {\bibinfo  {journal} {Phys. Lett. B}\ }\textbf {\bibinfo {volume} {550}},\
  \bibinfo {pages} {55} (\bibinfo {year} {2002})},\ \Eprint
  {https://arxiv.org/abs/hep-ph/0203134} {arXiv:hep-ph/0203134} \BibitemShut
  {NoStop}%
\bibitem [{\citenamefont {Kobayashi}\ and\ \citenamefont
  {Kugo}(1975)}]{Kobayashi:1975ev}%
  \BibitemOpen
  \bibfield  {author} {\bibinfo {author} {\bibfnamefont {M.}~\bibnamefont
  {Kobayashi}}\ and\ \bibinfo {author} {\bibfnamefont {T.}~\bibnamefont
  {Kugo}},\ }\href {https://doi.org/10.1143/PTP.54.1537} {\bibfield  {journal}
  {\bibinfo  {journal} {Prog. Theor. Phys.}\ }\textbf {\bibinfo {volume}
  {54}},\ \bibinfo {pages} {1537} (\bibinfo {year} {1975})}\BibitemShut
  {NoStop}%
\bibitem [{\citenamefont {Abbott}\ \emph {et~al.}(1976)\citenamefont {Abbott},
  \citenamefont {Kang},\ and\ \citenamefont {Schnitzer}}]{Abbott:1975bn}%
  \BibitemOpen
  \bibfield  {author} {\bibinfo {author} {\bibfnamefont {L.~F.}\ \bibnamefont
  {Abbott}}, \bibinfo {author} {\bibfnamefont {J.~S.}\ \bibnamefont {Kang}},\
  and\ \bibinfo {author} {\bibfnamefont {H.~J.}\ \bibnamefont {Schnitzer}},\
  }\href {https://doi.org/10.1103/PhysRevD.13.2212} {\bibfield  {journal}
  {\bibinfo  {journal} {Phys. Rev. D}\ }\textbf {\bibinfo {volume} {13}},\
  \bibinfo {pages} {2212} (\bibinfo {year} {1976})}\BibitemShut {NoStop}%
\bibitem [{\citenamefont {Linde}(1977)}]{Linde:1976qh}%
  \BibitemOpen
  \bibfield  {author} {\bibinfo {author} {\bibfnamefont {A.~D.}\ \bibnamefont
  {Linde}},\ }\href {https://doi.org/10.1016/0550-3213(77)90112-2} {\bibfield
  {journal} {\bibinfo  {journal} {Nucl. Phys. B}\ }\textbf {\bibinfo {volume}
  {125}},\ \bibinfo {pages} {369} (\bibinfo {year} {1977})}\BibitemShut
  {NoStop}%
\bibitem [{\citenamefont {Bardeen}\ and\ \citenamefont
  {Moshe}(1983)}]{Bardeen:1983st}%
  \BibitemOpen
  \bibfield  {author} {\bibinfo {author} {\bibfnamefont {W.~A.}\ \bibnamefont
  {Bardeen}}\ and\ \bibinfo {author} {\bibfnamefont {M.}~\bibnamefont
  {Moshe}},\ }\href {https://doi.org/10.1103/PhysRevD.28.1372} {\bibfield
  {journal} {\bibinfo  {journal} {Phys. Rev. D}\ }\textbf {\bibinfo {volume}
  {28}},\ \bibinfo {pages} {1372} (\bibinfo {year} {1983})}\BibitemShut
  {NoStop}%
\bibitem [{\citenamefont {Bardeen}\ and\ \citenamefont
  {Moshe}(1986)}]{Bardeen:1986td}%
  \BibitemOpen
  \bibfield  {author} {\bibinfo {author} {\bibfnamefont {W.~A.}\ \bibnamefont
  {Bardeen}}\ and\ \bibinfo {author} {\bibfnamefont {M.}~\bibnamefont
  {Moshe}},\ }\href {https://doi.org/10.1103/PhysRevD.34.1229} {\bibfield
  {journal} {\bibinfo  {journal} {Phys. Rev. D}\ }\textbf {\bibinfo {volume}
  {34}},\ \bibinfo {pages} {1229} (\bibinfo {year} {1986})}\BibitemShut
  {NoStop}%
\bibitem [{\citenamefont {Cao}\ \emph {et~al.}(2022)\citenamefont {Cao},
  \citenamefont {Li},\ and\ \citenamefont {Zheng}}]{Cao:2022zhn}%
  \BibitemOpen
  \bibfield  {author} {\bibinfo {author} {\bibfnamefont {X.-H.}\ \bibnamefont
  {Cao}}, \bibinfo {author} {\bibfnamefont {Q.-Z.}\ \bibnamefont {Li}},\ and\
  \bibinfo {author} {\bibfnamefont {H.-Q.}\ \bibnamefont {Zheng}},\ }\href
  {https://doi.org/10.1007/JHEP12(2022)073} {\bibfield  {journal} {\bibinfo
  {journal} {JHEP}\ }\textbf {\bibinfo {volume} {12}},\ \bibinfo {pages}
  {073}},\ \Eprint {https://arxiv.org/abs/2207.09743} {arXiv:2207.09743
  [hep-ph]} \BibitemShut {NoStop}%
\bibitem [{\citenamefont {Wang}\ \emph {et~al.}(2018)\citenamefont {Wang},
  \citenamefont {Yao},\ and\ \citenamefont {Zheng}}]{Wang:2017agd}%
  \BibitemOpen
  \bibfield  {author} {\bibinfo {author} {\bibfnamefont {Y.-F.}\ \bibnamefont
  {Wang}}, \bibinfo {author} {\bibfnamefont {D.-L.}\ \bibnamefont {Yao}},\ and\
  \bibinfo {author} {\bibfnamefont {H.-Q.}\ \bibnamefont {Zheng}},\ }\href
  {https://doi.org/10.1140/epjc/s10052-018-6024-5} {\bibfield  {journal}
  {\bibinfo  {journal} {Eur. Phys. J. C}\ }\textbf {\bibinfo {volume} {78}},\
  \bibinfo {pages} {543} (\bibinfo {year} {2018})},\ \Eprint
  {https://arxiv.org/abs/1712.09257} {arXiv:1712.09257 [hep-ph]} \BibitemShut
  {NoStop}%
\bibitem [{\citenamefont {Hoferichter}\ \emph {et~al.}(2023)\citenamefont
  {Hoferichter}, \citenamefont {de~Elvira}, \citenamefont {Kubis},\ and\
  \citenamefont {Mei\ss{}ner}}]{Hoferichter:2023mgy}%
  \BibitemOpen
  \bibfield  {author} {\bibinfo {author} {\bibfnamefont {M.}~\bibnamefont
  {Hoferichter}}, \bibinfo {author} {\bibfnamefont {J.~R.}\ \bibnamefont
  {de~Elvira}}, \bibinfo {author} {\bibfnamefont {B.}~\bibnamefont {Kubis}},\
  and\ \bibinfo {author} {\bibfnamefont {U.-G.}\ \bibnamefont {Mei\ss{}ner}},\
  }\href@noop {} {\  (\bibinfo {year} {2023})},\ \Eprint
  {https://arxiv.org/abs/2312.15015} {arXiv:2312.15015 [hep-ph]} \BibitemShut
  {NoStop}%
\end{thebibliography}%

\end{document}